\documentclass[onecolumn,11pt]{IEEEtran}

\usepackage{fullpage}

\usepackage{mathpazo}
\usepackage{times}
\usepackage{graphicx}
\usepackage{wrapfig}
\usepackage{epsfig}
\usepackage{graphicx}
\usepackage{color}
\usepackage{mdframed}
\usepackage{mathrsfs} 
\usepackage{amsfonts}
\usepackage{latexsym}
\usepackage{amsmath}
\usepackage{amssymb}
\usepackage{color}
\usepackage{float} 
\usepackage{floatflt} 
\usepackage{caption}
\usepackage{subcaption}
\usepackage{balance}
\usepackage{cite}
\usepackage{pdfpages}
\usepackage{setspace}
\usepackage{upref}
\usepackage{theorem}
\usepackage{hyperref}

\newcommand{\suppress}[1]{}

\newtheorem{theorem}{Theorem}[section]
\newtheorem{question}{Question}
\newtheorem{lemma}{Lemma}[section]

\newtheorem{claim}{Claim}[section]
\newtheorem{protocol}{Protocol}[section]

\newtheorem{definition}{Definition}[section]
\newtheorem{corollary}{Corollary}[section]

\newcommand{\rs}{{\tt sec}}
\newcommand{\rk}{{\tt key}}

\def\cX{\mbox{$\cal{X}$}}

\def\cR{\mbox{$\cal{R}$}}
\def\cB{\mbox{$\cal{B}$}}

\def\cK{\mbox{$\cal{K}$}}

\def\cZ{\mbox{$\cal{Z}$}}

\newcommand{\cI}{{{\cal I}}}

\newcommand{\bR}{{{\bf R}}}

\def\01{\{0,1\}}

\newcommand{\remove}[1]{}

\begin{document}

\title{Characterizing positive-rate key-cast \\(and multicast network coding) \\ with eavesdropping nodes}

\author{Michael Langberg\ \ \ \ \ \ \ \ \  \ \ \ \ \ \ \ \ \ 
Michelle Effros
\thanks{M. Langberg is with the Department of Electrical Engineering at the University at Buffalo (State University of New York).  
Email: \texttt{mikel@buffalo.edu}
M. Effros is with the Department of Electrical Engineering at the California Institute of Technology.
Email: \texttt{effros@caltech.edu}
This work is supported in part by NSF grant CCF-2245204.
Parts of this work appear in preliminary form in \cite{LE:24}.}}
%


\maketitle

\begin{abstract}
In multi-source multi-terminal key-dissemination, here called ``key-cast,'' introduced by the authors in [ITW2022], network nodes hold independent random bits, and one seeks a  communication scheme that allows all terminal nodes to share a secret key $K$.
The work at hand addresses positive (albeit, arbitrarily small) rate key-cast under the security requirement that no single non-terminal network node can gain information about the shared key $K$; this scenario is useful in cryptographic settings.  
Specifically, key-dissemination protocols based on secure multicast network coding are designed.
The analysis presented yields two combinatorial characterizations. 
In each, we assume a network in which an eavesdropper may access any individual network node.
The first characterization captures all networks that support positive-rate secure multicast; computing the secure-multicast {\em capacity} in the setting studied is a known open problem.
The second characterizes  all networks that support positive-rate secure key-cast. 
\end{abstract}

\thispagestyle{empty}

\section{Introduction} 
\label{sec:intro}

The resource of shared secret randomness, i.e., a shared secret key, plays a fundamental role in the theory and practice of network communication systems; applications include  cryptographic encryption, randomized coding technologies, distributed computing, statistical inference, distributed learning, distributed authentication, identification,  local differential-privacy, and more (e.g., \cite{shannon1949communication,maurer1993secret,ahlswede1993common,ahlswede1998common,ahlswede1978elimination,csiszar1988capacity,csiszar1991capacity,lapidoth1998reliable,KN:06,konevcny2016federated,alistarh2017qsgd,mugunthan2019smpai,acharya2020distributed,shlezinger2020federated,szabo2022optimal,chen2021communication,acharya2022role,lim2005extracting,su2008digital,suh2007physical,yu2009towards,acharya2019communication,byrd2020differentially,ahlswede2021identification}).
Motivated by the central role of shared randomness in such a wide range of distributed applications, the work at hand addresses the problem of disseminating common randomness over noiseless networks, for example, in the context of Network Coding \cite{ahlswede2000network,li2003linear,koetter2003algebraic,jaggi2005polynomial,ho2006random}; we call this the {\em key-cast} problem.
In the key-cast problem, introduced by the authors in \cite{langberg2022network}, network nodes hold independent random bits, and we seek a communication scheme that allows all terminal nodes to share a secret key $K$.


In this work, we focus on the cryptographically-motivated setting of key-cast in which one is only required to disseminate a {\em positive-rate} key, which, once shared among a collection of terminals, can be used to generate long sequences of common pseudo-random bits; the pseudo-random bits, in turn, can be used in applications like those mentioned above.
Our interest lies in {\em secret} key dissemination under a natural secrecy condition in which the shared key $K$ is independent of the information available at any non-terminal network node. 
As a result, in our setting, no network node other than the terminal nodes themselves, not even the nodes where random bits originate, learn any information about the secret key $K$ as they participate in communicating $K$ to the terminal nodes.

This work designs secure key-cast schemes that are strongly based on the notion of {\em secure multicast}.
In secure multicast one seeks to securely communicate source information to a collection of terminals in the presence of an eavesdropper with predefined eavesdropping capabilities.
The model of secure multicast network coding includes source nodes, that have access to message information, and additional nodes that generate independent randomness used to enable secure communication.
Most prior works on secure multicast, e.g., \cite{cai2002secure,feldman2004capacity,cai2007security,yeung2008optimality,el2012secure,silva2011universal,jaggi2012secure},  consider a single source setting in which the source $s$ generates both source messages and independent randomness, while no other network nodes can generate randomness. They further apply a {\em uniform} security assumption in which
the eavesdropper can access any collection of at most $z$ unit-capacity network links for a given security parameter $z$.
A major result in this context characterizes the secure multicast capacity and demonstrates that the capacity can be efficiently obtained using linear codes \cite{cai2002secure,yeung2008optimality,feldman2004capacity,el2012secure,silva2011universal}.
A more general model of secure multicast, where several network nodes can generate messages and/or independent randomness and eavesdroppers have access to  edge sets  with varying capacities (e.g., the setting of eavesdropping on nodes) is studied in, e.g., \cite{4460828,chan2014network,cui2012secure,chaudhuri2021characterization,huang2018,chaudhuri2018trade, chaudhuri2019secure};
in this general setting, the capacity is not fully characterized.
In fact, determining its value is known, in certain cases, to be NP-hard \cite{cui2012secure} or as hard as determining the capacity of the $k$-unicast problem \cite{huang2018} (a well known open problem in the study of network codes, e.g., \cite{yeung2006network,chan2014network,langberg2009multiple}). 

Our study focuses on the design of positive-rate key-cast schemes that are resilient against non-uniform eavesdroppers that can access the information available at any single node;
single-node access is considered to be non-uniform since each node may have a different in-degree.
Our scheme builds on a corresponding  positive-rate secure-multicast scheme in the setting in which any network node can generate randomness or messages and under the security requirement that no single internal network node can gain information about the transmitted message(s).
Towards that end, in Section~\ref{sec:main1}, we focus on the task of secure multicast and answer the following question (stated roughly below, and with greater rigor in Section~\ref{sec:model}) leading to a combinatorial characterization of networks that support positive-rate multicast resilient to eavesdroppers that control individual network nodes. 

\begin{question}[Positive-rate secure multicast]
\label{q:intro}
Given a communication network $G$ in which any network node can generate independent randomness, and given a set of terminal nodes $D$, is it possible to securely multicast a message $m$ from a potential source $s$ to  each node in $D$ such that no non-terminal network node (except $s$) gain information about $m$?
\end{question}

In Section~\ref{sec:key}, we focus on the task of secure key-cast and answer the following question (stated roughly below, and with greater rigor in Section~\ref{sec:model}) leading to a combinatorial characterization of networks that support positive-rate key-cast resilient to eavesdroppers that control individual network nodes. 

\begin{question}[Positive-rate secure key-cast]
\label{q:intro2}
Given a communication network $G$ in which any network node can generate independent randomness, and given a set of terminal nodes $D$, is it possible to securely disseminate a key $K$ to nodes in $D$ such that no non-terminal network node gain information about $K$?
\end{question}

\begin{figure}[t]
\begin{center}
\hspace*{25mm}  \includegraphics[width=1\columnwidth]{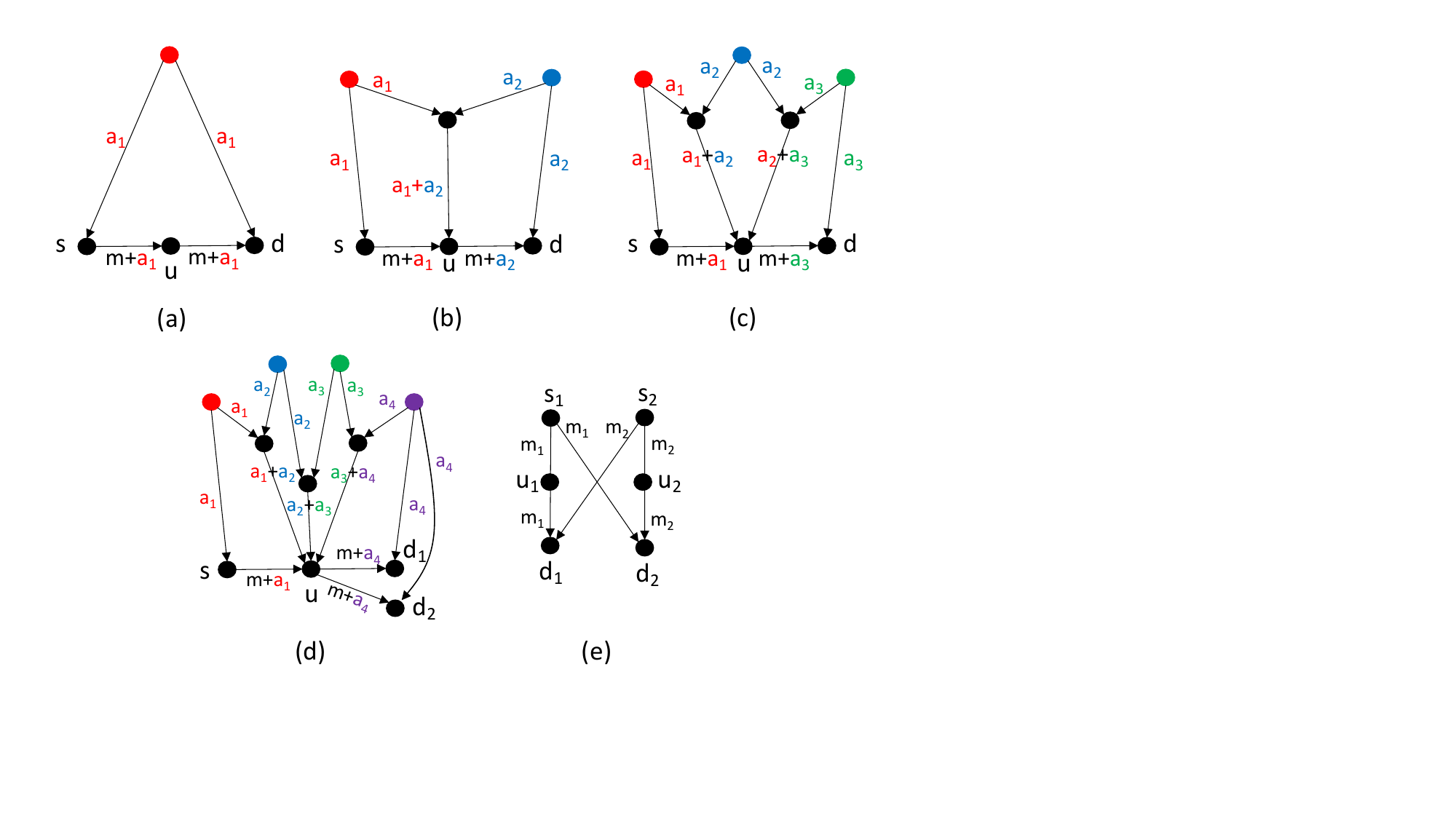}
\vspace{-18mm}
\caption{A number of examples corresponding to Questions~\ref{q:intro} and \ref{q:intro2} highlighting the major ideas used in our combinatorial characterization of networks that allow positive-rate secure multicast (Section~\ref{sec:main1}) and those allowing  positive-rate secure key-cast (Section~ \ref{sec:key}).
}
\label{fig:intro}
\end{center}
\end{figure}

Figure~\ref{fig:intro} depicts a number of examples corresponding to Questions~\ref{q:intro} and \ref{q:intro2}. 
In each example, any node can generate independently and uniformly distributed random bits.
In what follows, we review the examples in  Figure~\ref{fig:intro}, highlighting the major ideas used in our answers to 
Questions~\ref{q:intro} and \ref{q:intro2}.

\noindent
{\bf $\bullet$ Secure multicast.}
The networks depicted in Figures~\ref{fig:intro}(a)-(d) all allow the secure communication of message $m$ from source $s$ to terminal $d$, and thus exhibit cases for which secure communication is possible (i.e., cases in which  the answer to Question~\ref{q:intro} is positive).
In Figure~\ref{fig:intro}(a), the vertex $u$ is a {\bf cut vertex} that separates $s$ and $d$. 
Naively, one may assume that $u$ has the capabilities to gain information on any message transmitted between $s$ and $d$.
However, as noticed in prior works on secure network coding, e.g., \cite{4460828,chan2014network,cui2012secure,chaudhuri2021characterization,huang2018,chaudhuri2018trade, chaudhuri2019secure}, the information traversing the cut vertex $u$ can at times be {\em protected} using (a collection of) one-time pads.
We refer to each such vertex $u$ as a {\bf protected cut-vertex}.
Cut vertices (and protected cut-vertices) play a major role in our analysis in Sections~\ref{sec:main1} and \ref{sec:key}; see Definitions~\ref{def:cut} and \ref{def:protected2}.
Indeed, in Figure~\ref{fig:intro}(a) the red node can generate a uniformly distributed bit $a_1$ that is independent of $m$. As this node is connected to both $s$ and $d$, a one-time pad is established and $u$ does not gain information on $m$, implying secure communication.

The padding protocols protecting the information traversing $u$ may be more advanced than that of Figure\ref{fig:intro}(a). 
Additional examples are given in Figures\ref{fig:intro}(b)-(d).
The nodes colored in red, blue, green, and purple generate uniformly distributed and independent bits $a_1,\dots,a_4$, respectively, and through certain connectivity requirements, related to the notion of {\bf alternating paths} (see, Definition~\ref{def:alternating}), allow protection of the source information $m$ traversing $u$.
For example, in Figure\ref{fig:intro}(c), node $u$ receives $m+a_1$, $a_1+a_2$, and $a_2+a_3$.
Thus, $u$ can compute $m+a_3$ and send it to $d$ which has access to $a_3$ for decoding. 
In this process, $u$ does not gain information about $m$.
We formally define the connectivity requirements allowing the protection of information traversing $u$ and the corresponding {\bf ``padding'' protocol} in Definition~\ref{def:protected2} and Protocol~\ref{def:protocol}, respectively.
The characterization of networks for which the answer to Question~\ref{q:intro} is positive is given in Theorem~\ref{the:q1}.

\noindent
{\bf $\bullet$ Secure key-cast.}
The examples depicted in Figures~\ref{fig:intro}(a)-(d) also allow secure key-cast.
Recall that for secure key-cast no non-terminal network node (including the sources) gain any information on the shared key $K$.
In the case of a single terminal $d$, this is trivial, since $d$ can trivially generate its own key $K$.
But even in the case of Figure~\ref{fig:intro}(d), in which $D=\{d_1,d_2\}$, one can establish a shared key $K$
by sending an additional uniformly distributed and independent bit $m'$ from $u$ to $d_1$ and $d_2$.
This allows the terminals access to $m$ and $m'$, and accordingly to the key $K=m+m'$ which is independent of the information available at any non-terminal network node.
We note that while it is not always the case that networks allowing secure multicast (when $|D| >1$) also allow secure key-cast, not much is needed (with respect to the network topology) to convert a secure multicast scheme into a secure key-cast one. We elaborate on such extended schemes in Theorem~\ref{the:q1b}.

While secure multicast is a great first step towards a secure key-cast scheme, it is not hard to come up with examples in which secure key-cast is possible but secure multicast is not. 
One such example is given in Figures~\ref{fig:intro}(e).
Secure multicast from $s_1$ is not possible in this case since the cut vertex $u_1$ separating $s_1$ and $d_1$ is {\em unprotected}; similarly for $s_2$, the cut vertex $u_2$ separating $s_2$ and $d_2$ is unprotected.
However, for key-cast, one can send $m_1$ and $m_2$ to both $d_1$ and $d_2$, implying that they may share the key $K=m_1+m_2$ unknown to any other non-terminal node. 
Roughly speaking, while there exist distinct unprotected cut vertices separating source-terminal pairs $(s_1,d_1)$ and $(s_2,d_2)$,  the existence of a key-cast scheme follows from the fact that there is no {\em single} unprotected cut-vertex $u$ that separates {\em both} pairs.
We extend these ideas through formal definitions and statements in Section~\ref{sec:key}, culminating in the characterization of networks for which the answer to Question~\ref{q:intro2} is positive (Theorem~\ref{the:q1c} and Corollary~\ref{cor:q1c}).

The remainder of our presentation is structured as follows.
In Section~\ref{sec:model}, we present our detailed model and formalize Questions~\ref{q:intro} and \ref{q:intro2}.
The combinatorial characterization of networks that allow positive-rate secure multicast (i.e., for which the answer to Question~\ref{q:intro} is ``yes''),  together with a corresponding scalar-linear binary scheme, is given in Section~\ref{sec:main1}.
The combinatorial characterization of networks that allow positive-rate secure key-cast (i.e., for which the answer to Question~\ref{q:intro2} is ``yes''),  together with a corresponding scalar-linear binary scheme, is given in Section~\ref{sec:key}.


\section{Model}
\label{sec:model}

We follow the notation of \cite{langberg2022network}, modified here to address the positive-rate setting.
For any $\ell>0$, $[\ell]\triangleq\{1,2,\dots,\lceil \ell \rceil\}$.
\vspace{2mm}

\noindent
{\bf $\bullet$ Key-cast Instance:} An instance ${\mathcal I}=(G,S,D,\cB)$ of the key-cast problem includes an acyclic
directed network $G=(V,E)$,
a collection of source nodes $S \subseteq V$, a collection of terminal nodes $D \subseteq V$, and a collection $\cB=\{\beta_{1},\dots,\beta_{{\tiny |\cB|}}\}$ of subsets of edges  specifying the secrecy requirements.
Each source node $s_i \in S$ holds an unlimited collection $M_i=\{b_{ij}\}_j$ of independent, uniformly distributed bits.   
Let $M=\cup_{s_i \in S}M_i$ denote all random bits available at the source nodes.
Following a convention common in the study of acyclic network coding, we assume that 
the terminals $d \in D$ have no outgoing edges. 

\vspace{2mm}

\noindent
{\bf $\bullet$ Key-Codes:}
A network code $({\mathcal F},\mathcal{G})=(\{f_{e}\},\{g_{i}\})$, here called a {\em key-code},
is an assignment of an alphabet $\cX_e$ and a (local) encoding function
$f_{e}$ for each edge $e\in E$ and a decoding function $g_{i}$ for each terminal $d_{i} \in D$.
For every edge $e=(u,v)$, the edge message
$X_{e} \in \cX_{e}$ from $u$ to $v$ equals the evaluation of encoding function $f_{e}$  on inputs $X_{{\tt In}(u)}$;
here, for a generic node $u_0$, 
 $\text{In-edges}(u_0)$ is the collection $({e}:e = (v,u_0) \in E)$  of edges incoming to $u_0$, and  
$$X_{{\tt In}(u_0)} = ((X_e : e \in \text{In-edges}(u_0)), (\{b_{ij}\}_j: u_0=s_i))$$
captures all information available to node $u_0$ during the communication process.
Similarly, $\text{In-nodes}(u_0)$ equals $(v \in V:(v,u_0)\in E)$ is the collection of nodes $v$ with outgoing edges entering $u_0$.
In order to ensure that $X_{{\tt In}(u)}$ is available to node $u$ before it encodes, communication proceeds according to a predetermined topological order on $E$.

A key-code with target rate $R>0$ is considered successful if, for every terminal $d_{i} \in D$, the evaluation of decoding function $g_{i}$ on the vector of random variables $X_{{\tt In}(d_{i})}$ equals the reproduction of a uniform random variable $K$ over alphabet $\cK=[2^{R}]=\{1,2,\dots,\lceil 2^R \rceil\}$ such that the following criteria are satisfied.
First, key $K$ meets secrecy constraints $\cB$, which specifies that $I(K;(X_e: e \in \beta))=0$ for every $\beta \in \cB$.
Second, each terminal $d_i$ decodes key $K$.
Notice that the alphabets $\cX_e$ chosen in code design may be set to be arbitrarily large.
We thus refer to the setting at hand as ``positive-rate" since the rate per time step resulting from choosing a large alphabet size $\cX_e$ may be very small but still greater than zero.

\begin{definition}[Secure key-cast feasibility]
\label{def:key}
Instance $\cI$ is said to have positive key-cast rate $R_\rk>0$ if there exists a key-code $({\mathcal F},\mathcal{G})$ such that
\begin{itemize}
	\item {\bf Key Rate:} $K$ is a uniform random variable over $[2^{R_\rk}]$.
	\item {\bf Decoding:} For all $d_i \in D$, $H(K|X_{{\rm In}(d_i)})=0$.
	\item {\bf Secrecy:}  $I(K;(X_{e}:e \in \beta))=0$ for any subset $\beta \in \cB$.
\end{itemize}
\end{definition}
\vspace{2mm}


\noindent
{\bf $\bullet$ Secure multicast:} In the secure-multicast setting, one distinguishes between source nodes $S_m$ that hold message information and source nodes $S_r$ that hold independent randomness used for masking. The two subsets may intersect. As before, we assume that every node $s_i$ in $S_m \cup S_r$ holds an unlimited collection of independent bits $\{b_{ij}\}_j$. 

\begin{definition}[Secure-multicast feasibility]
\label{def:secure_mul}
Instance $\cI=(G,(S_m,S_r),D,\cB)$ is said to have positive secure-multicast rate $R_\rs>0$ if there exists a network code $({\mathcal F},\mathcal{G})$ such that
\begin{itemize}
	\item {\bf Message Rate:} $K$ is a uniform random variable over $[2^{R_\rs}]$ such that $K=M' \subset \{b_{ij}\}_{i \in S_m, j}$, where $M'$ is a subset of the source-bits generated by sources in $S_m$.
	\item {\bf Decoding:} For all $d_i \in D$, $H(K|X_{{\rm In}(d_i)})=0$.
	\item {\bf Secrecy:}  $I(K;(X_{e}:e \in \beta))=0$ for any subset $\beta \in \cB$.
\end{itemize}
\end{definition}

Notice that in both Definition~~\ref{def:key} and Definition~\ref{def:secure_mul}, the random variable $K$ is shared between the terminals in $D$. 
In the key-cast setting (\ref{def:key}), $K$ denotes the secret key, which may be a (uniformly distributed) function of source bits; the source bits themselves are not necessarily decoded at terminals in $D$.
In the secure multicast setting (\ref{def:secure_mul}), $K$ denotes the secret message generated at  sources in $S_m$ and decoded at each terminal in $D$.
It is thus evident that the task of key-cast is more flexible than that of secure multicast: roughly speaking, instance $\cI$ has positive key-cast rate $R_\rk>0$ according to Definition~\ref{def:key} if $\cI$ has positive  secure-multicast rate $R_\rs>0$ according to Definition~\ref{def:secure_mul}, 
but $R_\rk>0$ in Definition~\ref{def:key} does not ensure $R_\rs>0$ in Definition~\ref{def:secure_mul} since $R_\rk>0$ does not ensure decodability of even a single bit from $S_m$.

The work at hand addresses instances in which each network node can generate uniformly distributed independent random bits, i.e., the setting where $S=V$ in  Definition~\ref{def:key} and $S_r=V$ in Definition~\ref{def:secure_mul}.
Moreover, we consider eavesdroppers that have access to any individual network node (except terminal nodes).
Namely, in Definition~\ref{def:key} we consider 
$\cB=\{\beta_v \mid v \in V \setminus  D,  \beta_v=\text{In-edges}(v)\}$.\footnote{The security requirement expressed by $\cB$ implies that $I(K;(X_e: e \in \text{In-edges}(v)))=0$. Notice, by the definition of $X_{\tt In}(v)$, that this implies $I(K;X_{\tt In}(v))=I(K;(X_e: e \in \text{In-edges}(v)),M_i)=0$ as well; here taking into account the independent bits $M_i$ generated at $v$. That is, $K$ is independent of all information available to $v$.}
Similarly, in Definition~\ref{def:secure_mul} we require the message to be kept secret from any non-terminal node excluding message-generating sources.

In Section~\ref{sec:main1}, we seek to combinatorially characterize instances $\cI$ with positive secure-multicast rate.
We note that if there exists a secure-multicast scheme over $\cI=(G,(S_m,S_r),D,\cB)$ communicating positive rate $K$  with $|S_m|>1$, then there exists a positive-rate  {\em single message-source} secure-multicast scheme over $\cI=(G,(\{s\},S_r),D,\cB)$ for each $s \in S_m$ that generates message-bits in $K$.
We can construct the latter code from the former by replacing all random message bits in $K$ generated by nodes in $S_m\setminus \{s\}$ by constants.
We thus, without loss of generality, consider instances in which $S_m=\{s\}$. 
We seek to answer the following question, which formalizes Question~\ref{q:intro} from the Introduction.
\vspace{2mm}

\noindent
{\bf Question 1 (Positive-rate secure multicast)}
{\em For which instances  $\cI=(G,(\{s\},V),D,\cB)$ with $\cB=\{\beta_v \mid v \in V \setminus  (D \cup \{s\}),  \beta_v=\text{In-edges}(v)\}$ can we achieve $R_\rs>0$?}
\vspace{2mm}

In Section~\ref{sec:key}, we seek to combinatorially characterize instances $\cI$ with positive secure key-rate.
We thus seek to answer the following question, which formalizes Question~\ref{q:intro2} from the Introduction.
\vspace{2mm}

\noindent
{\bf Question 2 (Positive-rate secure key-cast)}
{\em For which instances  $\cI=(G,V,D,\cB)$ with $\cB=\{\beta_v \mid v \in V \setminus  D,  \beta_v=\text{In-edges}(v)\}$ can we achieve $R_\rk>0$?}
\vspace{2mm}

%
%

\section{Answering Question~\ref{q:intro}}
\label{sec:main1}

\begin{figure}[t]
\begin{center}
\hspace*{25mm}\includegraphics[width=1\columnwidth]{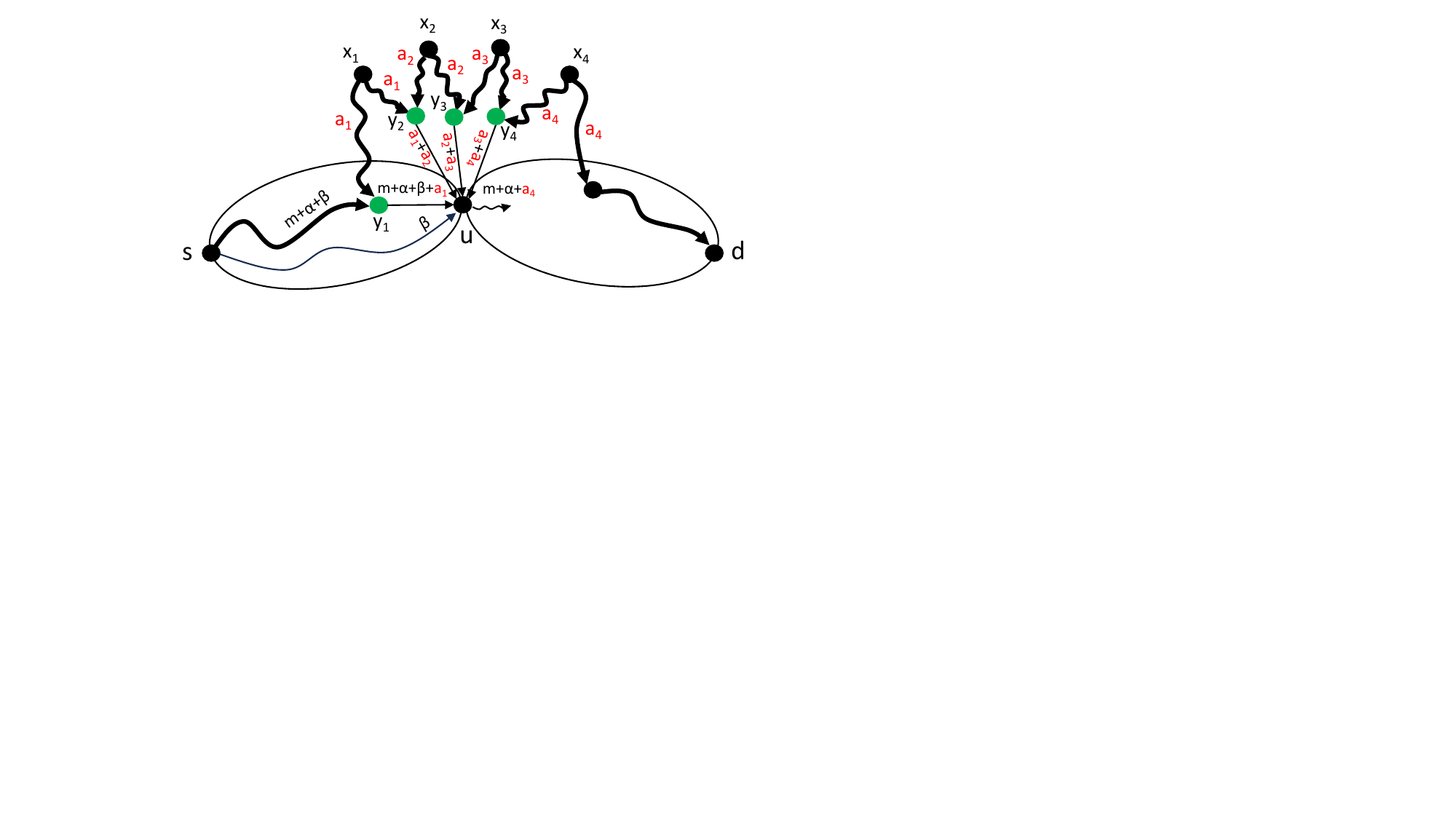}
\vspace{-60mm}
\caption{Depiction of Protocol~\ref{def:protocol} and Claim~\ref{claim:padding} for the case $\ell=4$.
The alternating path $P^{({\tt alt})}(s,d)$ is in bold.
Vertices $x_1,\dots,x_4$ generate independent random bits $a_1,\dots,a_4$ (in red) respectively.
The colliders $y_1,\dots,y_4$  (in green) along $P^{({\tt alt})}(s,d)$ mask the information traversing $u$.
Node $u$ is able to compute the linear combination $m+\alpha+a_\ell$ and does not gain any information about $m$.
The two large ovals represent the graph vertices that are reachable from $s$.}
\label{fig:q1}
\end{center}
\end{figure}

In this section we consider secure multicast instances $\cI=(G,(S_m=\{s\},S_r=V),D,\cB)$ in which $\cB=\{\beta_v \mid v \in V \setminus  (D \cup \{s\}),  \beta_v=\text{In-edges}(v)\}$.
That is, in $\cI$, we require the information multicast from $s$ to $D$ to be independent of the information available at any network node except the source $s$ and terminals in $D$.
We first consider the case where $|D|=1$, i.e., $D=\{d\}$ and analyze secure communication from $s$ to $d$.
We then address general $D$.
We start with a number of definitions and then describe a subroutine to be used in our analysis.
The definitions and subroutine are illustrated by 
Figure~\ref{fig:Wpath} and Figure~\ref{fig:q1}.

\subsection{Preliminary notation and definitions ($D=\{d\}$)}


\begin{definition}[Cut vertex]
\label{def:cut}
A vertex $u \in V$ is called a {\em cut vertex} for the source-terminal pair $(s,d)$ if the removal of $u$ separates $s$ from $d$ in $G$.
Equivalently, all paths from $s$ to $d$ go through $u$.
\end{definition}

\begin{figure}[t]
\begin{center}
\hspace*{25mm}  \includegraphics[width=1\columnwidth]{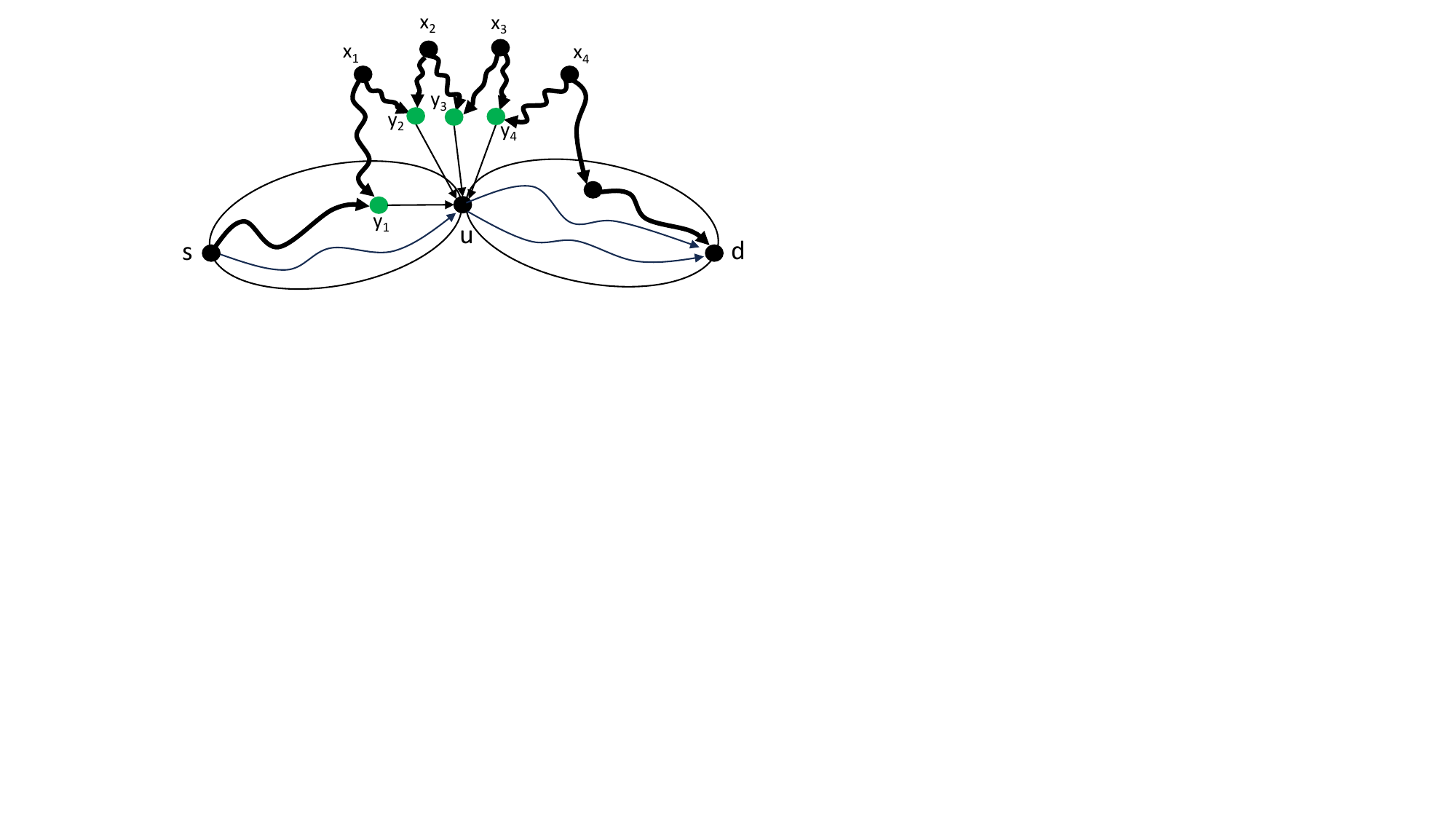}
\vspace{-60mm}
\caption{A depiction of an alternating path $P^{({\tt alt})}(s,d)$ (Definition~\ref{def:alternating}) and a protected cut vertex  $u$ separating $s$ and $d$ (Definition~\ref{def:protected2}).
The path $P^{({\tt alt})}(s,d)$ is given in bold and collider vertices $y_1,\dots,y_4$ on  $P^{({\tt alt})}(s,d)$ are colored in green.
Notice that all colliders are in $\text{In-nodes}(u)$.
The two large ovals represent the graph vertices that are reachable from $s$.}
\label{fig:Wpath}
\end{center}
\end{figure}

\begin{definition}[Alternating path]
\label{def:alternating}
Given a directed graph $G$ with vertex set $V$ and edge set $E$, 
its undirected variant, denoted by $\bar{G}$, has vertex set $V$ and undirected edge set $\bar{E}=\{(u,v) \mid (u,v) \in E\}$.
Nodes $v_1$ and $v_2$ are connected by an {\em alternating path} in $G$ if $v_1$ and $v_2$ are connected by a path $\bar{P}(v_1,v_2)$ in $\bar{G}$.
The alternating path $P^{({\tt alt})}(v_1,v_2)$ in $G$ connecting $v_1$ and $v_2$ consists of the set of edges $\{e\in E\mid \bar{e}\in \bar{P}\}$.
If vertex $v$ has two consecutive incoming edges in $P^{({\tt alt})}(v_1,v_2)$, then $v$ is called a {\em collider} vertex of  $P^{({\tt alt})}(v_1,v_2)$.
More specifically, $P^{({\tt alt})}(v_1,v_2)$ is an alternating path in $G$ if, for $\ell \geq 1$, there exist vertices $x_1,\dots,x_\ell$ and colliders $y_1,\dots,y_{\ell}$, such that $P^{({\tt alt})}(v_1,v_2)$ can be partitioned into path $P(v_1,y_1)$, paths $P(x_k,y_k)$ for $k \in [\ell]$, paths  $P(x_{k},y_{k+1})$ for $k \in [\ell-1]$, and path $P(x_{\ell},v_2)$;
see Figure~\ref{fig:Wpath} for an example. 
The paths above need not be disjoint; in addition, we may take {\em empty} paths of the form $P(v,v)$ for a vertex $v$.
\end{definition}

We next define the notion of ``protected cut-vertices.''
Roughly speaking, a cut vertex $u$ is protected if 
random variables generated at network nodes can be used to {\em mask } the message $m$ transmitted by $s$, thereby preventing $u$ from learning anything about $m$.

\begin{definition}[Protected cut vertices]
\label{def:protected2}
A cut vertex $u$ with respect to $(s,d)$ is {\em protected} if there exists an alternating path $P^{({\tt alt})}(s,d)$ that does not include $u$ and in addition all colliders $v$ on $P^{({\tt alt})}(s,d)$ are in $\text{In-nodes}(u)$.
See Figure~\ref{fig:Wpath}.
\end{definition}

\subsection{Statement of main theorem for Section~\ref{sec:key}}
We are now ready to state the main theorem for this section.

\begin{theorem}
\label{the:q1}
Let $\cI=(G,(S_m=\{s\},S_r=V),D=\{d\},\cB)$ with $\cB=\{\beta_v \mid v \in V \setminus (D \cup \{s\}), \beta_v=\text{In-edges}(v)\}$.
Then $\cI$ has secure-multicast rate $R_\rs>0$ according to Definition~\ref{def:secure_mul} if and only if every cut vertex $u$ separating $s$ and $d$ is protected.
\end{theorem}

\subsection{The ``padding'' protocol}
Before addressing the proof of Theorem~\ref{the:q1}, given a protected cut-vertex $u$, we preset our padding protocol (depicted in Figure~\ref{fig:q1}) specifying how information can flow through the network to mask the message $m$ transmitted by the source node $s$.
The secure-multicast scheme suggested shortly to answer Question~\ref{q:intro} uses the padding protocol repeatedly.
Towards that end, in addition to the message $m$, the protocol assumes the existence of random variables $\alpha$ and $\beta$ generated throughout the secure-multicast scheme and available to certain network nodes.
The justification for these assumptions,  in addition to the precise information content of $\alpha$ and $\beta$, are presented later when the padding protocol is used in our secure-multicast scheme given in the proof of Theorem~\ref{the:q1}

\begin{protocol}[Padding protocol for protected $u$]
\label{def:protocol}
Let $u$ be a protected cut-vertex according to Definition~\ref{def:protected2}.
Consider the alternating path $P^{({\tt alt})}(s,d)$ of Definition~\ref{def:protected2}.
Specifically, for $\ell \geq 1$, there exist vertices $x_1,\dots,x_\ell$ and $y_1,\dots,y_{\ell}$ in $V$ such that $P^{({\tt alt})}(s,d)$ consists of path $P(s,y_1)$, paths $P(x_k,y_k)$ for $k \in [\ell]$, paths  $P(x_{k},y_{k+1})$ for $k \in [\ell-1]$, and path $P(x_{\ell},d)$; none of which pass through $u$.
Moreover, all colliders $y_1,\dots,y_\ell$ are in $\text{In-nodes}(u)$.

For $k \in [\ell]$, let $a_k$ be a uniform bit generated independently at $x_k$.
Let  $m, \alpha,\beta \in \{0,1\}$ be random variables such that the variables in the collection  $\{m, \alpha,\beta,a_{1},\dots,a_{\ell}\}$ are mutually independent. 

The protocol described below (and depicted in Figure~\ref{fig:q1}) assumes that
\begin{itemize}
\item  node $y_1$ has access to $m+\alpha+\beta$ (all additions are \hspace{-2mm}$\mod 2$),
\item  node $u$ has access to $\beta$, and
\item  node $u$ may or may-not have access to $\alpha$,
\end{itemize} 
and guarantees that 
\begin{itemize}
\item node $u$ can compute the linear combination $m+\alpha+a_\ell$.
\item assuming knowledge of $\alpha$ and $\beta$, node $u$ does not gain any information about $m$.
\end{itemize} 
The protocol proceeds as follows.
\begin{itemize}
\item  For $k \in [\ell-1]$, node $x_k$ sends $a_k$ to $y_k$ through $P(x_k,y_k)$ and to $y_{k+1}$ through $P(x_k,y_{k+1})$. 
\item Node $x_{\ell}$ sends $a_\ell$ to $d$ through $P(x_\ell,d)$ and to $y_\ell$ through $P(x_\ell,y_\ell)$.
\item Node $y_1$, with access to $m+\alpha+\beta$ and $a_1$,  sends $m+\alpha+\beta+a_1$ to $u$.
\item For $k =2, \dots, \ell$, node $y_k$ sends $a_{k-1}+a_{k}$ to $u$.
\end{itemize}
This concludes the padding protocol.
\end{protocol}

Under the given definitions, each of the nodes $x_1,\dots,x_\ell$ is the source of a one-time pad (here denoted by $a_1,\dots,a_{\ell}$) independent of all other random variables in the network.
These one-time pads are sequentially added and then removed from the source message to ensure that the source message remains protected.
Protection at the bottleneck $u$ is achieved by transmitting to the bottleneck not the protection bits themselves but sums of consecutive pairs of those bits.
Namely, we conclude the following claim.

\begin{claim}
\label{claim:padding}
In Protocol~\ref{def:protocol}, assuming that node $y_1$ has access to $m+\alpha+\beta$ and that 
node $u$ has access to $\beta$, it holds that (i) node $u$ is able to compute the linear combination $m+\alpha+a_\ell$, and (ii) assuming knowledge of $\alpha$ and $\beta$, node $u$ does not gain any information about $m$.
\end{claim}

\begin{proof}
Let $\ell \geq 1$.
For (i), the proof follows from the fact that 
node $u$, knowing $\beta$ and, using incoming information from $y_1,\dots,y_{\ell}$, can compute 
{\small{$
\beta+ (m+\alpha+\beta+a_1)+ (a_1+a_{2})+(a_{{2}}+a_{3})+  \dots + (a_{{\ell-1}}+a_{{\ell}})
=m+\alpha+ a_\ell.
$}}
For (ii), note that the incoming information to $u$ is independent of $m$; namely,
{\small{
$
I(\alpha,\beta,m+\alpha+\beta+a_1,a_1+a_{2},a_{{2}}+a_{3}, \dots ,a_{{\ell-1}}+a_{{\ell}};m)=0.
$}}
Thus, the information available at $u$ through Protocol~\ref{def:protocol} is independent of $m$ (even if $u$ knows $\alpha$). 
\end{proof}

\subsection{Proof of Theorem~\ref{the:q1}}
\label{sec:proof_q1}
We are now ready to prove the main result of this section.

\begin{proof}(of Theorem~\ref{the:q1})
We start by proving {\bf achievability}. 
%
If there are no cut vertices in $V$, then $s$ and $t$ are 2-vertex connected, meaning that there exist two vertex-disjoint paths, $P_1(s,d)$ and $P_2(s,d)$, in $G$ between $s$ and $d$ \cite{menger1927allgemeinen}.
Then $\bR_{\tt SR} \geq 1$
since, given independent uniformly distributed bits $m$ and  $r$, the source can send $m+r$ on $P_1$ and $r$ on $P_2$. 
The resulting scheme is secure.

\begin{figure}[t]
\begin{center}
\hspace*{25mm}\includegraphics[width=1\columnwidth]{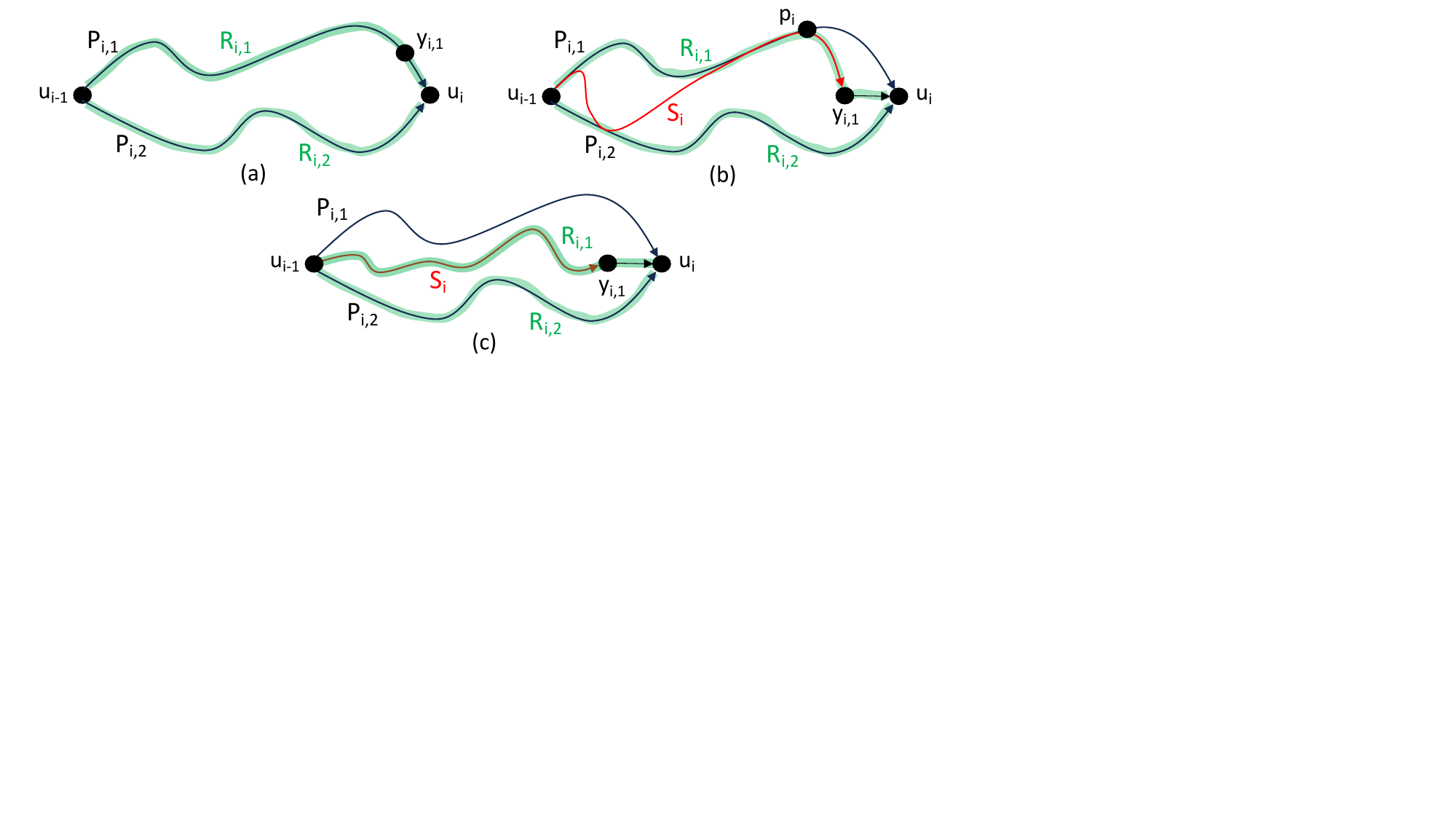}
\vspace{-55mm}
\caption{Depiction of the paths $P_{i,1}, P_{i,2}$ (in black), $R_{i,1}, R_{i,2}$ (in green) and $S_i$ (in red) from the proof of Theorem~\ref{the:q1}. In case (a),   $P_{i,1}$ includes the vertex $y_{i,1}\in \text{In-nodes}(u_i)$. In case (b), $S_i$ intersects the paths $P_{i,1}$ or $P_{i,2}$ (or both). In case (c),   $S_i$ does not intersect the paths $P_{i,1}$ or $P_{i,2}$. }
\label{fig:paths}
\end{center}
\end{figure}

Otherwise, let $u_1,\dots,u_c$ be the collection of cut-vertices ordered topologically, and let $U_{1}, \dots, U_{c+1}$ be the partition of $V$ implied by $u_1,\dots,u_c$ in which $U_{i}$ includes all vertices of $V$ of topological order less than that of $u_i$ that are not in $\cup_{j<i}U_{j}$. 
Denote the vertex $s$ by $u_0$ and the vertex $d$ by $u_{c+1}$.
It follows for every $i \in \{1,\dots,c+1\}$ that either $u_{i-1}$ and $u_{i}$ are connected by two vertex-disjoint paths $P_{i,1}(u_{i-1},u_{i})$ and $P_{i,2}(u_{i-1},u_{i})$ or that $(u_{i-1},u_{i}) \in E$.
The latter occurs if and only if $|U_{i}|=1$, which implies that {$U_{i}=\{u_{i-1}\}$} and that $u_{i-1}$ is the only incoming node to $u_i$ that is reachable from $s$.
We refer to sets $U_{i}$ of size 1 as type-1 sets; and other all $U_{i}$ as type-2 sets.

Before proposing the secure communication scheme, we set additional notation.
Recall that we assume each vertex $u_i$ for $i \in [c]$ to be protected. 
Thus, for each $i \in [c]$, by Protocol~\ref{def:protocol} 
there exists an alternating path $P^{({\tt alt})}_i(s,d)$ including vertices $x_{i,1},\dots,x_{i,\ell_i}$,\ $y_{i,1},\dots,y_{i,\ell_i}$ and corresponding paths 
$P(s,y_{i,1})$,  $P(x_{i,k},y_{i,k})$ for $k \in [\ell_i]$,   $P(x_{i,k},y_{i,k+1})$ for $k \in [\ell_i-1]$, and  $P(x_{\ell_i},d)$; none of these paths pass through $u$.
Moreover, all colliders $y_{i,1},\dots,y_{i,\ell_i}$ are in $\text{In-nodes}(u_i)$.
In addition, by Claim~\ref{claim:padding}, using Protocol~\ref{def:protocol}, and assuming that $u_i$ receives information $m+\alpha_i+\beta_i+a_{i,1}$  and $\beta_i$
for $a_{i,1}$ generated at $x_{i,1}$, then $u_i$ can compute $m + \alpha_i+a_{i,\ell_i}$ in a way that keeps all incoming information available to $u_i$ collectively independent of $m$. 

For $i \in [c+1]$, if $|U_{i}|>1$, recall the existence of two vertex-disjoint paths $P_{i,1}(u_{i-1},u_i)$ and $P_{i,2}(u_{i-1},u_i)$ from $u_{i-1}$ to $u_i$. 
We next define two additional vertex-disjoint paths $R_{i,1}(u_{i-1},u_i)$ and  $R_{i,2}(u_{i-1},u_i)$, depicted in Figure~\ref{fig:paths},  
such that $R_{i,1}$ passes through $y_{i,1}$.
If $y_{i,1}$ is on $P_{i,1}$ (or $P_{i,2}$), see Figure~\ref{fig:paths}(a), then take $R_{i,1}$ and $R_{i,2}$ to be equal to $P_{i,1}$ and $P_{i,2}$ (or equal to $P_{i,2}$ and $P_{i,1}$) respectively.
Otherwise, let $S_i(u_{i-1},y_{i,1})$ be a path from $u_{i-1}$ to $y_{i,1}$.
Such a path must exist, since otherwise there is no path from $s$ to $y_{i,1}$ (as guaranteed by the fact that $u_i$ is protected).
If  $S_{i}$ intersects $P_{i,1}$ or $P_{i,2}$ (see Figure~\ref{fig:paths}(b)), let $p_i$ be the vertex of maximum topological order in the intersection.
Assume, without loss of generality, that $p_i$ is on $P_{i,1}$.
Set $R_{i,1}$ to be equal to $P_{i,1}$ until $p_i$ and then equal to $S_{i}$ followed by the edge $(y_{i,1},u_i)$.
Let $R_{i,2}$ equal $P_{i,2}$.
It holds that $R_{i,1}$ and $R_{i,2}$ are vertex-disjoint.
Finally, in the case where $y_{i,1}$ is not on $P_{i,1}$ or $P_{i,2}$ and also $S_{i}$ does not intersect $P_{i,1}$ or $P_{i,2}$ (see Figure~\ref{fig:paths}(c)),
 set $R_{i,1}$ equal to $S_i$ followed by the edge $(y_{i,1},u_i)$ and $R_{i,2}$ equal to $P_{i,2}$.
Again, it holds that $R_{i,1}$ and $R_{i,2}$ are vertex-disjoint.


We are now ready to suggest a secure communication scheme in which the source $s$ securely sends a uniform bit $m$ to $d$.
The scheme is described sequentially over the sets $U_{i}$ for $i=1,\dots,c+1$ and shows by induction that, for $i \in [c]$, $u_i$ is able to communicate $m+\sum_{j=1}^{i}a_{j,\ell_j}$ without learning anything about $m$.

If $U_{1}$ is type-1, then it must be the case that $s$ is the only incoming vertex to $u_1$ that is reachable from $s$ and thus $y_{1,1}=s$.
Using the notation of Protocol~\ref{def:protocol} with constant $\alpha_1=\beta_1 \equiv 0$, it holds that $y_{1,1}=s$ has access to $m+\alpha_1+\beta_1=m$ and $u_1$ has access to $\beta_1$.
Thus applying Protocol~\ref{def:protocol} on $u_1$ guarantees that $u_1$ can compute $m+\alpha_1+a_{1,\ell_1}=m+a_{1,\ell_1}$ without gaining information about $m$.
Vertex $u_1$ prepares to send $m+a_{1,\ell_1}$ in the next step of communication.

\begin{figure}[t]
\begin{center}
\hspace*{25mm}\includegraphics[width=1\columnwidth]{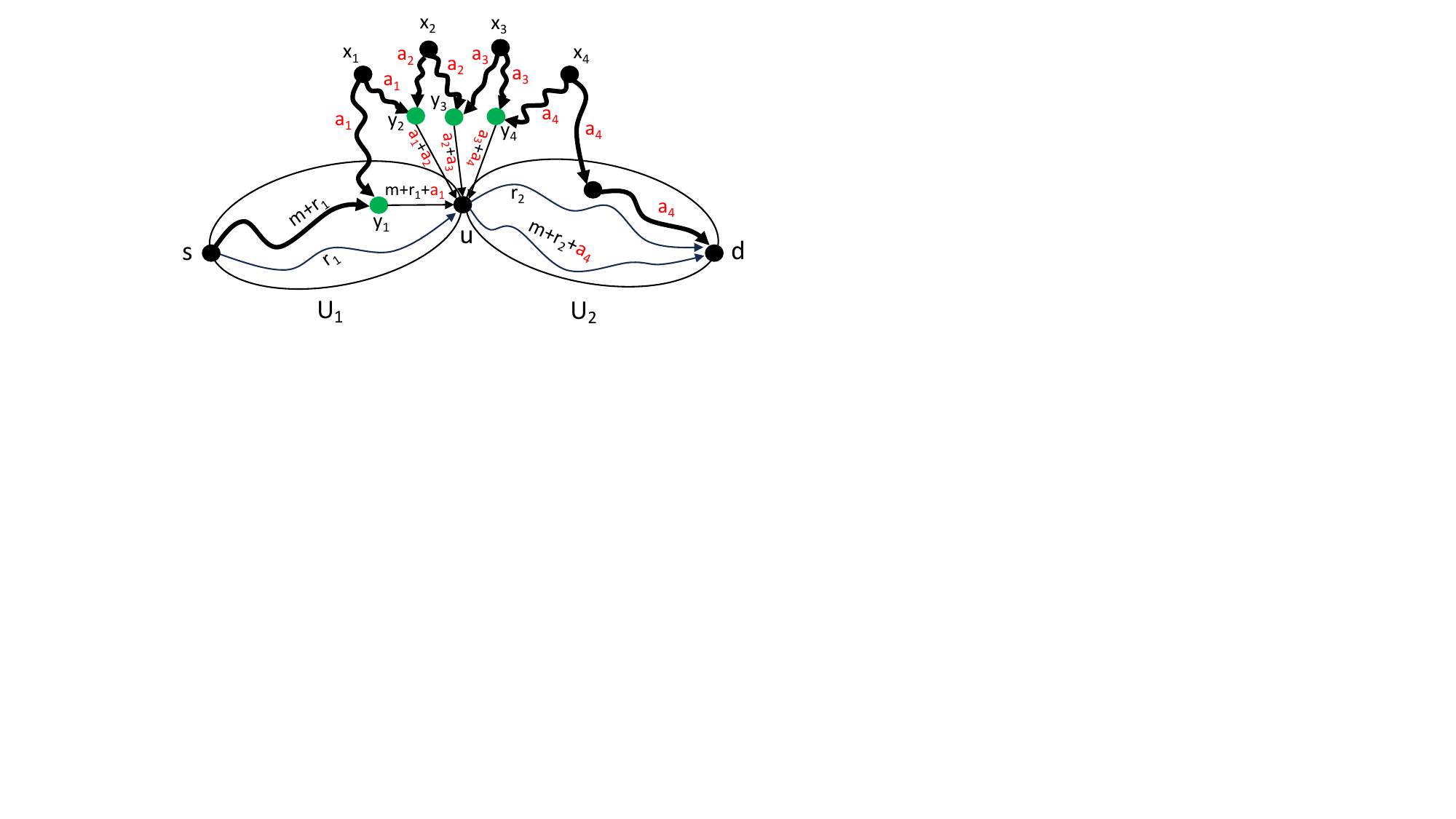}
\vspace{-55mm}
\caption{An example for the achievability of Theorem~\ref{the:q1} in which $c=1$. The two large ovals represent the graph vertices that are reachable from $s$.}
\label{fig:example}
\end{center}
\end{figure}

If $U_{1}$ is type-2, then $s$ sends $m+r_1$ on path $R_{1,1}$ until $y_{1,1}$ and $r_1$ on  path $R_{1,2}$, where $r_1$ is an independent uniformly distributed bit.
Now, using the notation of Protocol~\ref{def:protocol} with constant $\alpha_1\equiv 0$ and $\beta_1=r_1$, it holds that $y_{1,1}$ has access to $m+\alpha_1+\beta_1=m+r_1$ and $u_1$ has access to $\beta_1=r_1$.
Thus, applying Protocol~\ref{def:protocol} on $u_1$ guarantees that $u_1$ can compute $m+\alpha_1+a_{1,\ell_1}=m+a_{1,\ell_1}$ without gaining information about $m$.
Notice that no other node in $U_{1} \setminus \{s\}$ gains information about $m$.
%
%
%
Vertex $u_1$ prepares to send $m+a_{1,\ell_1}$ in the next step of communication.

The protocol for an intermediate $U_{i}$ follows the previous analysis with $\alpha_i$ of Protocol~\ref{def:protocol}  set to $\sum_{j=1}^{i-1}a_{j,\ell_j}$. We here assume by induction that $u_{i-1}$ has computed and is prepared to transmit $m+\sum_{j=1}^{i-1}a_{j,\ell_j}$. 
Specifically, 
if $U_{i}$ is type-1, then it must be the case that $y_{i,1}= u_{i-1}$, i.e., $u_{i-1}$ is the only incoming vertex to $u_i$ that is reachable from $s$.
Thus, using the notation of Protocol~\ref{def:protocol} with $\alpha_i=\sum_{j=1}^{i-1}a_{j,\ell_j}$, and $\beta_i \equiv 0$, it holds that $y_{i,1}=u_{i-1}$ has access to $m+\alpha_i+\beta_i=m+\sum_{j=1}^{i-1}a_{j,\ell_j}$ and $u_i$ has access to $\beta_i$.
Thus applying Protocol~\ref{def:protocol} on $u_i$ guarantees that $u_i$ can compute $m+\alpha_i+a_{i,\ell_i}=m+\sum_{j=1}^{i}a_{j,\ell_j}$ without gaining information on $m$ (even under the assumption that $u_i$ knows $\alpha_i=\sum_{j=1}^{i-1}a_{j,\ell_j}$ from previous rounds of communication).

If $U_{i}$ is type-2, then $u_{i-1}$ sends $m+\sum_{j=1}^{i-1}a_{j,\ell_j}+r_i$ on path $R_{i,1}$ until $y_{i,1}$ and $r_i$ on  path $R_{i,2}$, where $r_i$ is an independent uniformly distributed bit.
Now, using the notation of Protocol~\ref{def:protocol} with $\alpha_i = \sum_{j=1}^{i-1}a_{j,\ell_j}$ and $\beta_i=r_i$, it holds that $y_{i,1}$ has access to $m+\alpha_i+\beta_i=m+\sum_{j=1}^{i-1}a_{j,\ell_j}+r_i$ and $u_i$ has access to $\beta_i=r_i$.
Thus, applying Protocol~\ref{def:protocol} on $u_i$ guarantees that $u_i$ can compute $m+\alpha_i+a_{i,\ell_i}=m+\sum_{j=1}^{i}a_{j,\ell_j}$ without gaining information on $m$.
Notice that all other nodes in $U_{i}$ do not gain information on $m$ as well.
The above holds even under the assumption that $u_i$ knows $\alpha_i=\sum_{j=1}^{i-1}a_{j,\ell_j}$ from previous rounds of communication. 

For decoding, consider $U_{c+1}$.
For $U_{c+1}$ of type-1, $u_c$ forwards $m+\sum_{j=1}^{c}a_{j,\ell_j}$ to $d=u_{c+1}$.
As $d$ has access to $\{a_{{1,\ell_1}},\dots,a_{{c,\ell_c}}\}$, it can decode $m$.
For $U_{c+1}$ of type-2, $u_c$ sends $m+\sum_{j=1}^{c}a_{j,\ell_j}+r_c$ for a uniform $r_c$ on path $P_{c+1,1}(u_c,d)$ and $r_c$ on  path $P_{c+1,2}(u_c,d)$.
It follows that no node in $U_{c+1}$ gains information about $m$ (even if that node knows $\sum_{j=1}^{c}a_{j,\ell_j}$).
Since $d$ has access to $\{a_{{1,\ell_1}},\dots,a_{{c,\ell_c}}\}$, it can decode $m$.
This concludes the achievability proof.
Figure~\ref{fig:example} shows an example of the scheme for $c=1$.

We now address the {\bf converse.} 
Let $u$ be a cut vertex separating $s$ and $d$.
Suppose  that $u$ is not protected.
We show that there does not exist a positive-rate secure scheme that communicates a message $m$ from $s$ to $d$.
Assume in contradiction that $s$ is able to communicate $m$ securely to $d$.
Let $M$ be the random variable corresponding to $m$. 
The proof relies on Lemma~\ref{claim:converse} below, which employs the following definitions.
%
%
Let $W_u$ be the collection of vertices $v \in V$ for which there exists a path $P(v,u)$ in $G$.
Let $W_{s,u}$ be the set of vertices $v \in W_u$ for which there exists an alternating path $P^{({\tt alt})}(s,v)$ in $G$ that does not pass through $u$ and for which all colliders are in $\text{In-nodes}(u)$.
Let $\bar{W}_{s,u} = W_u \setminus W_{s,u}$.
See Figure~\ref{fig:converse}.


\begin{lemma}
\label{claim:converse}
Let $u$ be an unprotected cut vertex with respect to $(s,d)$.
Let $W_u$,  $W_{s,u}$, $\bar{W}_{s,u}$ be defined as above.
Then, (i) $s \in {W}_{s,u}$, (ii) any path $P(v,d)$ for $v \in {W}_{s,u}$ must pass through $u$,  and (iii) there are no edges in $G$ from vertices in ${W}_{s,u}$ to those in $\bar{W}_{s,u}$ or vice-versa.
Namely,
$$
E \cap ((W_{s,u}  \times \bar{W}_{s,u})  \cup (\bar{W}_{s,u} \times W_{s,u} ) ) =\phi.
$$
\end{lemma}

\begin{proof}
For (i), note that by the definition of $W_{s,u}$ using the empty path $P^{({\tt alt})}(s,s)$, it holds that $s \in W_{s,u}$.
For (ii), let $v \in {W}_{s,u}$. Then there exists an alternating path
$P^{({\tt alt})}(s,v)$ from $s$ to $v$ for which all colliders are in $\text{In-nodes}(u)$.
Assume in contradiction the existence of a path $P(v,d)$ that does not pass through $u$.
Concatenating the  alternating path $P^{({\tt alt})}(s,v)$ with $P(v,d)$, we obtain an alternating path $P^{({\tt alt})}(s,d)$ that does not pass through $u$ and for which all colliders are in $\text{In-nodes}(u)$.
The latter implies by Definition~\ref{def:protected2} that $u$ is protected, which contradicts the assumption that $u$ is not protected.

Finally, for (iii), we consider two cases.
If $\bar{W}_{s,u}=\phi$, then the assertion holds trivially.
Otherwise, both $W_{s,u}$ and $\bar{W}_{s,u}$ are non-empty.
Let $v \in W_{s,u}$ and $\bar{v} \in \bar{W}_{s,u}$.
See Figure~\ref{fig:converse}.
Assume, in contradiction, the existence of an edge $(v,\bar{v})$ or $(\bar{v},v)$ in $G$.
This implies an alternating path $P^{({\tt alt})}(s,\bar{v})$ from $s$ to $\bar{v}$ that does not pass through $u$ and for which all colliders (except perhaps $v$) are in $\text{In-node}(u)$.
If node $v$ in  $P^{({\tt alt})}(s,\bar{v})$ is not a collider, or alternatively, if node $v$ is a collider and $v \in \text{In-node}(u)$, then we have established that $\bar{v} \in  W_{s,u}$; this contradicts our assumption that $\bar{v} \in  \bar{W}_{s,u}$.
If node $v$ in $P^{({\tt alt})}(s,\bar{v})$ is a collider and $v \not \in \text{In-node}(u)$, then we slightly modify  $P^{({\tt alt})}(s,\bar{v})$ by (twice) adding a path $P(v,y)$ for $y \in \text{In-node}(u)$.
The latter is possible since $v \in W_u$.
The resulting modified alternating path again establishes that $\bar{v} \in  W_{s,u}$, which  contradicts our assumption that $\bar{v} \in  \bar{W}_{s,u}$.
\end{proof}

We are ready to complete the converse proof.
Let ${R}$ be the information generated at vertices in $W_{s,u}$.
Let $\bar{R}$ be the information generated at vertices in $\bar{W}_{s,u}$.
Notice that ${R}$ is independent of $\bar{R}$ as $W_{s,u} \cap \bar{W}_{s,u} = \phi$.
The information received by vertex $u$ through incoming vertices in  $W_{s,u}$ is a function of $R$;
let $f(R)$ represent all information generated by vertices in $W_{s,u}$ and received by vertex $u$.
The information received by vertex $u$ through incoming vertices  in $\bar{W}_{s,u}$ is a function of $\bar{R}$;
let $\bar{f}(\bar{R})$ represent all information generated by vertices in $\bar{W}_{s,u}$ and received by vertex $u$.

\begin{figure}[t]
\begin{center}
\hspace*{25mm}\includegraphics[width=1\columnwidth]{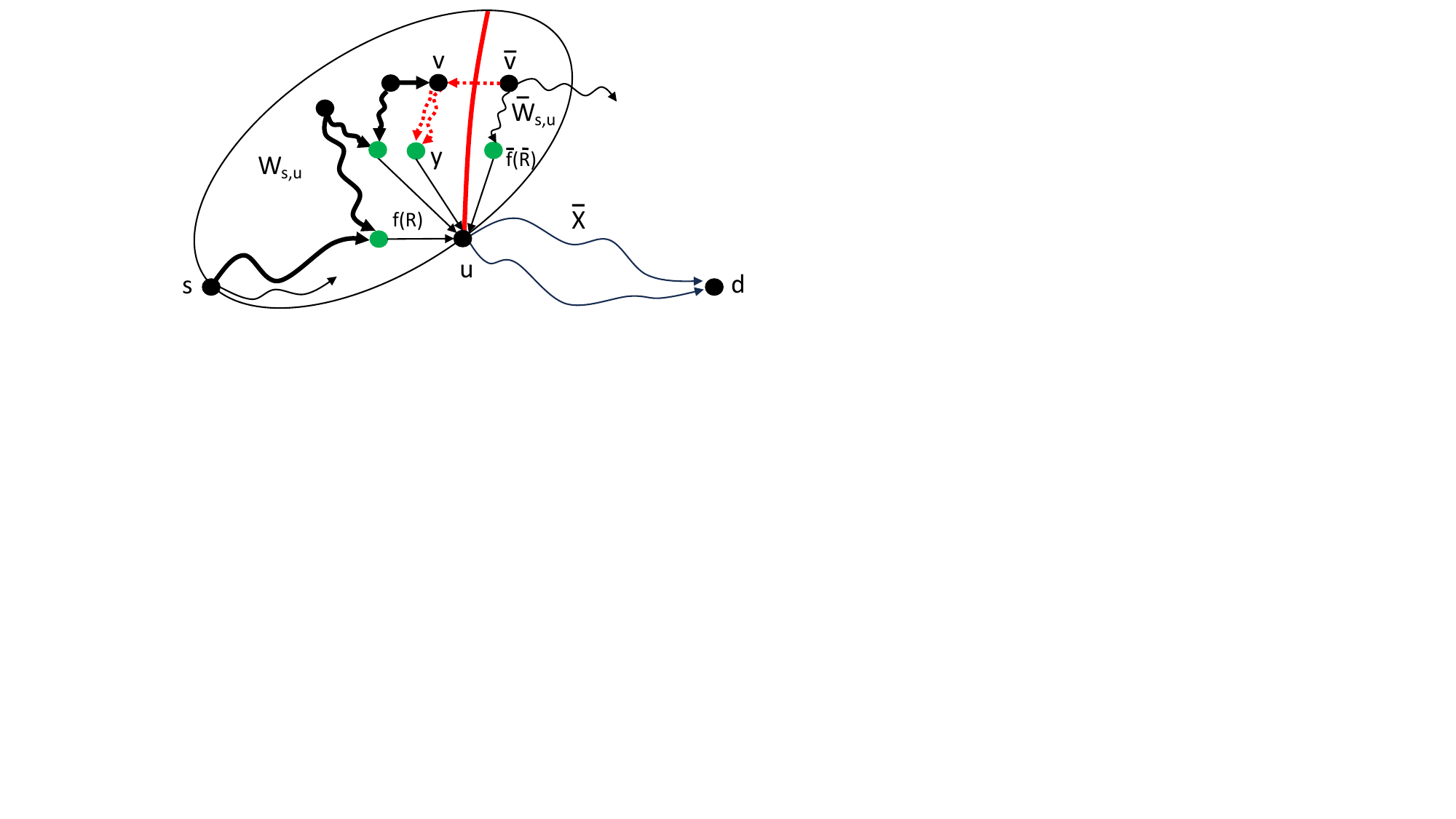}
\vspace{-55mm}
\caption{A depiction of the analysis in Lemma~\ref{claim:converse} and in the converse of Theorem~\ref{the:q1} (for cut vertex $u$). 
The large oval represents the set $W_u$ which is partitioned into $W_{s,u}$ and $\bar{W}_{s,u}$.
The red dotted edge $(\bar{v},v)$ represents the assumption made in-contradiction in the proof of part (iii) of Lemma~\ref{claim:converse}.
The red dotted path $P(v,y)$ is added (twice) to the alternating path $P^{({\tt alt})}(s,\bar{v})$ to obtain the contradiction.
The random variables $R$ and $\bar{R}$ are generated by vertices in $W_{s,u}$ and $\bar{W}_{s,u}$ respectively.
}
\label{fig:converse}
\end{center}
\end{figure}

The security requirement at $u$ is
$
I(M;f(R),\bar{f}(R))=0.
$
Since $u$ is a cut vertex, using Lemma~\ref{claim:converse}(ii), the  information incoming to $d$ is a function of the information at $u$ and additional randomness $\bar{X}$ generated at $V \setminus W_{s,u}$.
Thus, $(M,R)$ is independent of $(\bar{X},\bar{R})$.
Since $d$ can recover $M$, it holds that
$
I(M;f(R),\bar{f}(\bar{R}),\bar{X})>0.
$
However, we reach a contradiction by observing that
\begin{align*}
I(M;&f(R),\bar{f}(\bar{R}),\bar{X}) \\
& = 
I(M;f(R),\bar{f}(\bar{R}))  +
I(M;\bar{X} \mid f(R),\bar{f}(\bar{R})) \\
& \leq 
I(M,f(R);\bar{X},\bar{f}(\bar{R})) =0,
\end{align*}
where the last inequality follows since $I(A;B|C,D) \leq I(A,C;B,D)$ for any collection of random variables $A,B,C,$ and $D$.
This concludes the proof of Theorem~\ref{the:q1}.
\end{proof}

\subsection{Multiple terminals}
When $D$ includes more than one terminal node, we can perform the scheme described in Theorem~\ref{the:q1} for each terminal node $d_i$ in parallel with independent randomness; 
this yields a valid code since our positive-rate model allows arbitrary edge alphabets.
Formal details appear in Appendix~\ref{app:inter}.
The following corollary summarizes the result.
\begin{corollary}
\label{cor:q1}
Let $\cI=(G,(S_m=\{s\},S_r=V),D,\cB)$ with $\cB=\{\beta_v \mid v \in V \setminus (D \cup \{s\}), \beta_v=\text{In-edges}(v)\}$.
Then $\cI$ has secure-multicast rate $R_\rs>0$ according to Definition~\ref{def:secure_mul} if and only if, for every $d \in D$, every cut vertex $u$ with respect to $(s, d)$ is protected.
\end{corollary}

\section{Answering Question~\ref{q:intro2}}
\label{sec:key}

We show below that whenever the key-cast rate is positive, one can construct a key-cast scheme by slightly modifying the protocol of Theorem~\ref{the:q1};
this observation yields a combinatorial characterization for key-cast.
We start with a brief motivating discussion outlining  our methodology.

First assume that in $\cI$ there is a vertex $s$ such that one can securely multicast a (uniformly distributed) message bit $m$ to all terminals in $D$.
Then, by Corollary~\ref{cor:q1}, for every $d \in D$, every cut vertex $u$ separating the pair $(s,d)$ is protected. 
If, in addition, there is another node $s'$ that is connected to all terminals $d \in D$ in $G \setminus \{s\}$, then  one can  construct a secure key-cast scheme as follows. 
Source node $s$ securely multicasts a (uniformly distributed) message bit $m$ to all terminals in $D$. 
In parallel, source node $s'$ sends a (uniformly distributed, independent) bit $m'$ to all terminal nodes in $D$ along paths that do not pass through $s$.
Here, $m'$ is independent of all information used in the secure transmission of $m$, but its transmission need not be secure.
The key $K$ is set to $m+m'$.
One can now verify that the suggested scheme is secure (a full proof appears in Theorem~\ref{the:q1b}  in Appendix~\ref{app:keycast}). 

It is thus relatively simple to establish a key-cast scheme if from vertex $s$ one can securely multicast a (uniformly distributed) message bit $m$ to all terminals in $D$ using Corollary~\ref{cor:q1} (assuming additional connectivity requirements on $G$).
The challenge occurs when one wishes to establish a key-cast scheme in networks in which, for each node $s$, there exists a terminal $d \in D$ and a cut vertex $u$ separating the pair $(s,d)$ for which $u$ is not protected.
We call such $(s,d)$ pairs {\em unprotected}.
In what follows, we slightly modify the protocol of Theorem~\ref{the:q1} to handle unprotected $(s,d)$ pairs.
We show that while the modified protocol no longer supports secure multicast, it suffices as a building block for secure key-cast.


\subsection{Modified protocol for unprotected $(s,d)$ pairs}
\label{sec:modify}
Below, we use the notation of Section~\ref{sec:main1}.
For source $s$ and terminal $d$, let $u_1,\dots,u_c$ be the cut-vertices separating $(s,d)$;  $u_1,\dots,u_c$ are by assumption ordered topologically.
Let $U_{1}, \dots, U_{c+1}$ be the partition of $V$ implied by $u_1,\dots,u_c$.
\begin{protocol}[Modified protocol]
\label{protocol:2}
{\em
The modified protocol for unprotected $(s,d)$ pairs follows the protocol described in Theorem~\ref{the:q1} with the following changes:
\begin{itemize}
\item For each $U_{i}$ for which $u_i$ is protected, the protocol from $u_{i-1}$ to $u_i$ remains unchanged.
\item For each $U_{i}$ for which $u_i$ is not protected, node $u_{i-1}$ sends its outgoing information to $u_i$ (along the edge $(u_{i-1},u_i)$ for $U_i$ of type-1, or alternatively using the two paths $P_{i,1}(u_{i-1},u_i)$ and $P_{i,2}(u_{i-1},u_i)$ for $U_i$ of type-2) without invoking the padding protocol (Protocol~\ref{def:protocol}).
\end{itemize}}
\end{protocol}
In Protocol~\ref{protocol:2}, above, each protected cut-vertex $u_i$ gains no information about $m$, but each unprotected cut vertex $u_i$ may completely learn $m$.

\subsection{Statement of main theorem for Section~\ref{sec:key}}
We are now ready to state the main theorem for this section.
\begin{theorem}
\label{the:q1c}
Let $\cI=(G,V,D,\cB)$, with $\cB=\{\beta_v \mid v \in V \setminus D, \beta_v=\text{In-edges}(v)\}$.
Instance $\cI$ has key-cast rate $R_\rk>0$ according to Definition~\ref{def:key} if and only if there exists a subset $S=\{s_1,s_2,\dots,s_k\}$ of nodes such that running 
Protocol~\ref{protocol:2} for every $(s_i,d_j)$ pair in $S \times D$ (with message $m_i$ for source $s_i$)  disseminates the key $K=\sum_{i=1}^km_i$ securely to all terminals in $D$.
\end{theorem}

The proof of Theorem~\ref{the:q1c}, presented shortly, implies the following combinatorial characterization for positive secure key-rate.
\begin{corollary}
\label{cor:q1c}
Let $\cI=(G,V,D,\cB)$ in which $\cB=\{\beta_v \mid v \in V \setminus D, \beta_v=\text{In-edges}(v)\}$.
Instance $\cI$ has key-cast rate $R_\rk>0$ according to Definition~\ref{def:key} if and only if there exists a subset $S=\{s_1,s_2,\dots,s_k\}$ of nodes such that for every vertex $u \in V \setminus S$, the set 
\begin{align*}
S_u=\{s \in S \mid \exists \ d \in D\ \text{s.t.} \ \text{$u$ is an unprotected cut vertex} \ &\text{w.r.t. $(s,d)$} \}
\end{align*}
does not equal $S$,
and for every $s_i \in S$, 
the set 
\begin{align*}
S_{s_i}=\{s \in S \setminus \{s_i\} \mid \exists \ d \in D\ \text{s.t.} \ \text{$s_i$ is an unprotected cut vertex} \ &\text{w.r.t. $(s,d)$}\} 
\end{align*}
does not equal $S \setminus \{s_i\}$.
\end{corollary}

We start by presenting a number of definitions and claims to be used in our proofs.

\subsection{Preliminary definitions and claims}
\label{sec:pre}
Throughout this section, we consider a positive-rate key-cast scheme for a given instance $\cI=(G,V,D,\cB)$ in which $\cB=\{\beta_v \mid v \in V \setminus D, \beta_v=\text{In-edges}(v)\}$.
Let $M_v$ be the independent uniform bits generated in the scheme by node $v \in G$ and let $K$ be the shared key.
Define ${\tt Support}(K)$ as 
$$
{\tt Support}(K)\triangleq\{v \in V \mid H(K \mid (M_u: u \in V \setminus \{v\} )) >0\};
$$
that is, the key $K$ solely depends on the information generated at nodes $v \in {\tt Support}(K)$.  
We call the nodes in ${\tt Support}(K)$ {\em sources} and define ${\tt Support}(K)=\{s_1,\dots,s_k\}$.

In our analysis, we focus on the case in which there exists a single vertex $u$ that acts as an unprotected cut-vertex between $s_i$ and some $d_i \in D$ for every $s_i \in {\tt Support}(K)$.
We call such a vertex $u$ an {\em unprotected cut-vertex} for ${\tt Support}(K)$.
It holds that $u \in V \setminus {\tt Support}(K)$.
Consider any source terminal pair $(s_i,d_i)$ separated by the unprotected cut vertex $u$.
We first repeat a number of definitions from Section~\ref{sec:proof_q1}.
Let $W_u$ be the collection of vertices $v \in V$ for which there exists a path $P(v,u)$ in $G$.
Let $W_{s_i,u}$ consist  of all vertices $v \in W_u$ for which there exists an alternating path $P^{({\tt alt})}(s_i,v)$ in $G$ that does not pass through $u$ and for which all colliders are in $\text{In-nodes}(u)$.
Let $\bar{W}_{s_i,u} = W_u \setminus W_{s_i,u}$.
From Lemma~\ref{claim:converse} used in the proof of Theorem~\ref{the:q1}, we learn that 
(i) $s_i \in {W}_{s_i,u}$, (ii) any path $P(v,d)$ for $v \in {W}_{s_i,u}$ must pass through $u$,  and (iii) there are no edges in $G$ from vertices in $W_{s_i,u}$ to those in $\bar{W}_{s_i,u}$ or vice-versa, giving
$
E \cap ( (W_{s_i,u}  \times \bar{W}_{s_i,u})\cup (\bar{W}_{s_i,u} \times W_{s_i,u} )) =\phi.
$
We also note that ${\tt Support}(K) \subset W_u$ as $u$ is a cut vertex separating, for each $i \in [k]$, the pair $(s_i,d_i)$.

\begin{figure}[t]
\begin{center}
\hspace*{25mm}\includegraphics[width=1\columnwidth]{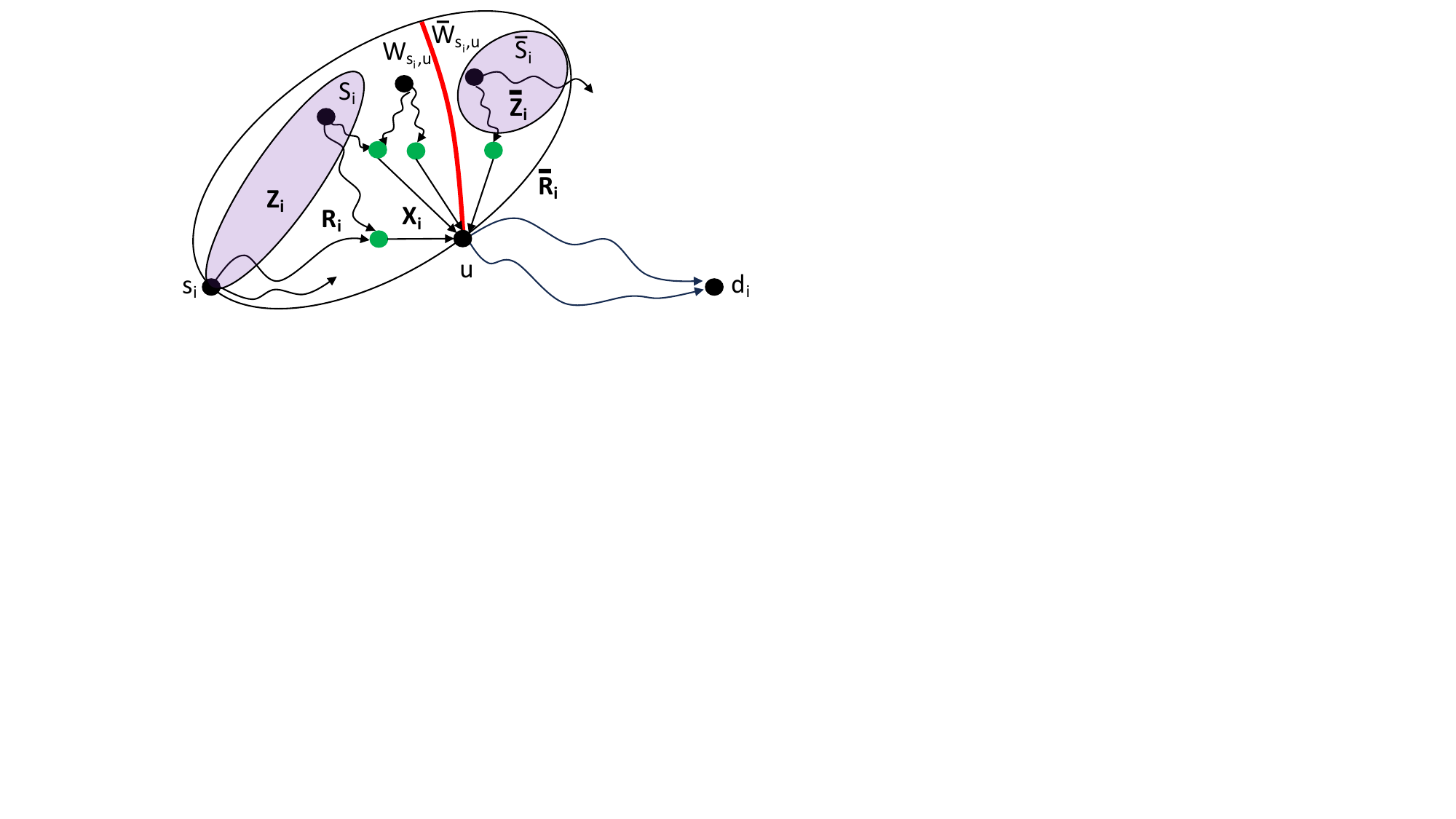}
\vspace{-55mm}
\caption{A depiction of the random variables $Z_i, R_i, \bar{Z}_i$, $\bar{R}_i$, and $X_i$ (in bold).
The large oval represents the set $W_u$ which is partitioned into $W_{s_i,u}$ and $\bar{W}_{s_i,u}$.
The sets $S_i$ and $\bar{S}_i$ are shaded in purple.
Any edge $e$ incoming to $u$ holds information $X_e$ for which either $H(X_e\mid Z_i,R_i)=0$ or $H(X_e\mid \bar{Z}_i,\bar{R}_i)=0$.
$X_i$ is the information carried on edges in $\text{In-edges}(u) \cap W_{s_i,u}$.
Incoming information to $d_i$, and thus the key $K$, is a function of $X_i$ and additional information that relies on $\bar{Z}_i$ and $\bar{R}_i$}
\label{fig:key}
\end{center}
\end{figure}

For every $s_i \in {\tt Support}(K)$, we define the following partition $(S_i,\bar{S}_i)$ of   ${\tt Support}(K)$:
$S_i = {\tt Support}(K) \cap W_{s_i,u}$ and $\bar{S}_i = {\tt Support}(K) \cap \bar{W}_{s_i,u}$.
Note that $s_i \in S_i$.
Here, both $S_i$ and $\bar{S}_i$ depend on cut vertex $u$, however, as the analysis that follows employs a fixed $u$, we simplify our presentation by using notation that does not explicitly highlight this dependence.  See Figure~\ref{fig:key}.


\begin{claim}
\label{claim:wpath1}
Let $u$ be an unprotected cut-vertex for ${\tt Support}(K)$.
Then, for any $s_i \ne s_j$ in ${\tt Support}(K)$, the corresponding sets $W_{s_i,u}$ and $W_{s_j,u}$ are either equal or disjoint.
\end{claim}
\begin{proof}
Suppose that there is some $v \in W_{s_i,u} \cap W_{s_j,u}$.
Then there exist alternating paths $P^{({\tt alt})}(s_i,v)$ and $P^{({\tt alt})}(s_j,v)$ that do not pass through $u$ and for which all colliders are in $\text{In-nodes}(u)$.
We show that  $W_{s_i,u} = W_{s_j,u}$.
To prove that $W_{s_i,u} \subseteq W_{s_j,u}$, consider $v_i \in W_{s_i,u}$.
Then there exists an alternating path $P^{({\tt alt})}(s_i,v_i)$ from $s_i$ to $v_i$ that does not pass through $u$ and for which all colliders are in $\text{In-nodes}(u)$.
Concatenating the alternating paths $P^{({\tt alt})}(s_j,v)$, $P^{({\tt alt})}(s_i,v)$, and $P^{({\tt alt})}(s_i,v_i)$ yields an alternating path $P^{({\tt alt})}(s_j,v_i)$ from $s_j$ to $v_i$ that starts at $s_j$, passes through $v$, passes through $s_i$, and concludes at $v_i$.
Similar to the analysis in Lemma~\ref{claim:converse}, the existence of $P^{({\tt alt})}(s_j,v_i)$ implies that $v_i \in W_{s_j,u}$.
Specifically, $P^{({\tt alt})}(s_j,{v}_i)$ does not pass through $u$ and has colliders (except perhaps $v$ and $s_i$) that are in $\text{In-node}(u)$.
If in  $P^{({\tt alt})}(s_j,{v}_i)$ nodes $v$ and $s_i$ are not colliders, or alternatively, if node $v$ ($s_i$) is a collider and $v \in \text{In-node}(u)$ ($s_i \in \text{In-node}(u)$), then we have established that  $v_i \in W_{s_j,u}$.
If in $P^{({\tt alt})}(s_j,{v}_i)$ node $v$ is a collider and $v \not \in \text{In-node}(u)$, then we slightly modify  $P^{({\tt alt})}(s_j,{v}_i)$ by (twice) adding a path $P(v,y)$ for $y \in \text{In-node}(u)$.
A similar modification is made if in $P^{({\tt alt})}(s_j,{v}_i)$ node $s_i$ is a collider and $s_i \not \in \text{In-node}(u)$.
Finally, a similar proof implies that  $W_{s_j,u} \subseteq W_{s_i,u}$, concluding our assertion.
\end{proof}

Combining Lemma~\ref{claim:converse}(i) with Claim~\ref{claim:wpath1} yields Claim~\ref{claim:wpath} stated below.
\begin{claim}
\label{claim:wpath}
Let $u$ be an unprotected cut-vertex for ${\tt Support}(K)$.
Then, for any $s_i \ne s_j$ in ${\tt Support}(K)$, if $s_j \in \bar{S}_i$, the sets $S_i$ and $S_j$ are disjoint.
\end{claim}

Considering the key cast scheme at hand,
for every $s_i \in {\tt Support}(K)$, let $Z_i$ be the information generated at nodes $S_i$, and let $\bar{Z}_i$ be the information generated at nodes $\bar{S}_i$.
Let $R_i$ be the information generated at nodes $W_{s_i,u} \setminus S_i$, and  let $\bar{R}_i$  be the information generated at nodes $V \setminus (W_{s_i,u} \cup \bar{S}_i)$.
The variables $Z_i, R_i, \bar{Z}_i$, and $\bar{R}_i$ determine the incoming information to $d_i$.
Moreover, the variables above are generated at disjoint subsets of vertices in $G$, and therefore  $Z_i, R_i, \bar{Z}_i$, and $\bar{R}_i$ are mutually independent.
See Figure~\ref{fig:key}.

Consider any $s_i \in {\tt Support}(K)$ and let $d_i \in D$ be a terminal for which $u$ is an unprotected cut-vertex for the pair $(s_i,d_i)$.
By Lemma~\ref{claim:converse}(ii), it holds that, for each source $s_j \in S_i$, any path from $s_j$ to terminal $d_i$ passes through $u$.
Moreover, as $u$ is unprotected, it follows from Lemma~\ref{claim:converse}(iii) that any edge $e$ incoming to $u$ holds information $X_e$ for which either $H(X_e\mid Z_i,R_i)=0$ or $H(X_e\mid \bar{Z}_i,\bar{R}_i)=0$; the former holds for edges $e \in \text{In-edges}(u) \cap W_{s_i,u}$ and the latter for the remaining incoming edges of $u$.
Let 
$X_i \triangleq X_{\text{In-edges}(u) \cap W_{s_i,u}}$ be the information carried on the set of edges 
$
\{e \in \text{In-edges}(u) \mid H(X_e\mid Z_i,R_i)=0\}.
$
We conclude that the incoming information to $d_i$ is a function of  $X_i$ and of  additional information that relies on $\bar{Z}_i$ and $\bar{R}_i$, and thus so too is the key $K$.
Thus, $H(X_i \mid Z_i,R_i)=0$ and there exists some function $f_1$ for which $K=f_1(X_i,\bar{Z}_i,\bar{R}_i)$. By our definition of ${\tt Support}(K)$ it is also true that there exists some function $g$ for which $K=g(Z_i,\bar{Z}_i)$. 
These observations are summarized in the following claim.
\begin{claim}
\label{claim:sum}
Consider any key-cast scheme for $\cI=(G,V,D,\cB)$.
Let ${\tt Support}(K)$ be defined as above, let $u$ be an unprotected cut vertex for ${\tt Support}(K)$, and let $s_i \in {\tt Support}(K)$ and $d_i \in D$ be a source-terminal pair for which $u$ is an unprotected cut vertex.
For $X_i$, $R_i$, $\bar{R}_i$, $Z_i$, and $\bar{Z}_i$ as defined above it holds that 
\begin{itemize}
\item $R_i$, $\bar{R}_i$, $Z_i$, and $\bar{Z}_i$ are independent.
\item Information $X_i$ is known to node $u$. 
\item $X_i$ is a function of $Z_i$ and $R_i$.
\item $K$ is a function of $Z_i$ and $\bar{Z}_i$.
\item $K$ is a function of $X_i$, $\bar{Z}_i$, and $\bar{R}_i$.
\end{itemize}
\end{claim}

\begin{figure}[t!]
\begin{center}
\hspace*{25mm}\includegraphics[width=1.25\columnwidth]{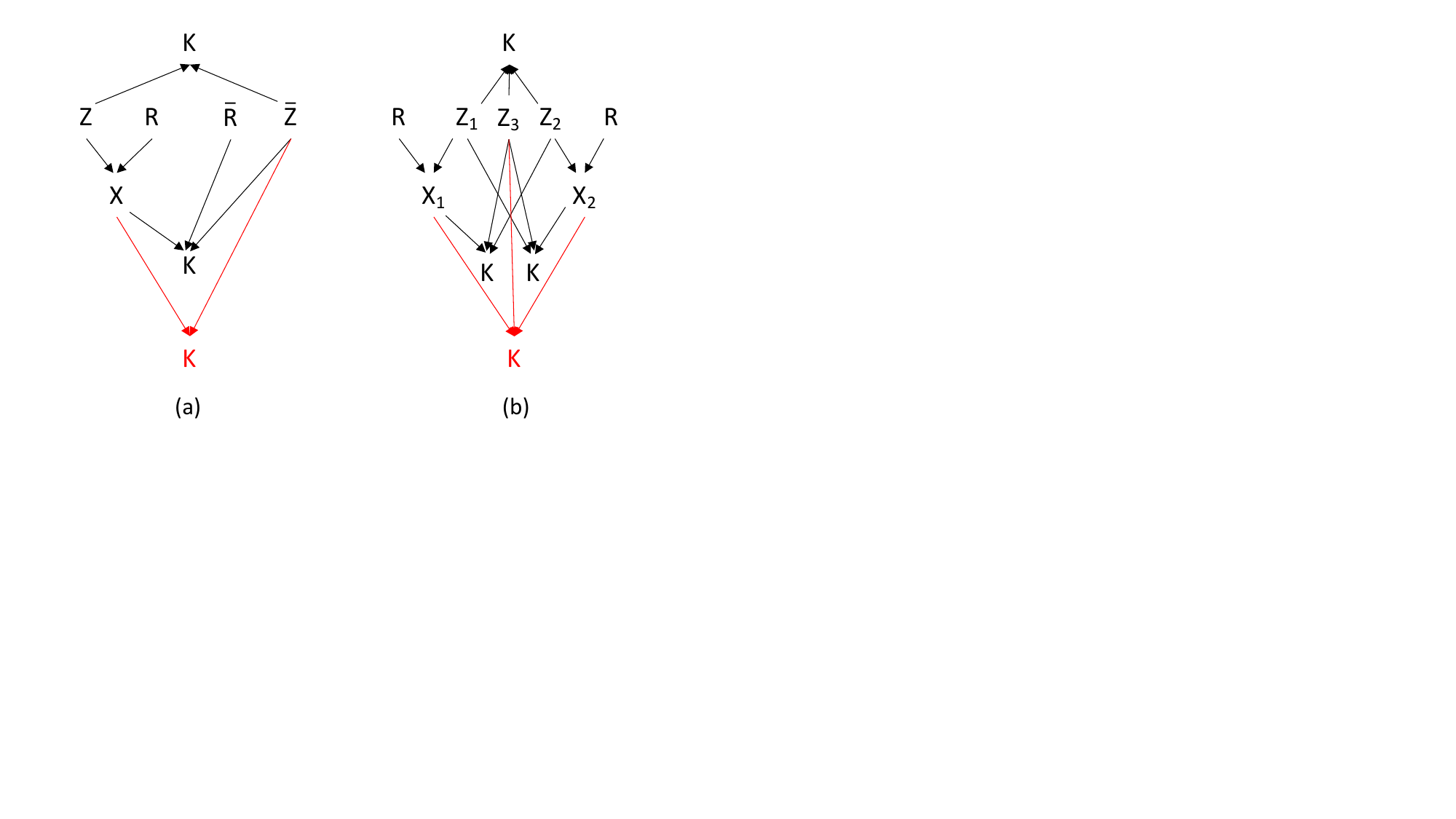}
\vspace{-58mm}
\caption{(a) Depiction of Claim~\ref{claim:func0}. The assumptions are depicted in black: $X$ is a function of $Z$ and $R$, $K$ is a function of $Z$ and $\bar{Z}$, and a function of $X$, $\bar{R}$, and $\bar{Z}$. The assertion is depicted in red:  $K$ is a function of $X$ and $\bar{Z}$. (b) Similarly, for Claim~\ref{claim:func}, with assumptions:
$X_1$ is a function of $R$ and $Z_1$, 
$X_2$ is a function of $Z_2$ and $R$,
$K$ is a function of $Z_1,Z_2,Z_{3}$, a function of $Z_1$, $Z_{3}$, $X_2$, and  a function of $X_1$, $Z_3$, $Z_{2}$.
The assertion (depicted in red) claims that $K$ is a function of $X_1$, $Z_3$, $X_2$. Note that $R$ is depicted twice in (b).}
\label{fig:dep}
\end{center}
\end{figure}

Our analysis employs the following information-processing claims, {illustrated in Figures~\ref{fig:dep}(a) and \ref{fig:dep}(b).}
\begin{claim}
\label{claim:func0}
Let $Z,\bar{Z},R,\bar{R}$ be independent random variables, and let $X$ be an additional random variable.
If,
\begin{itemize}
\item  $X=f_1(Z,R)$, 
\item  $K=g(Z,\bar{Z})$, and
\item  $K=g_1(X,\bar{Z},\bar{R})$
\end{itemize}
for functions $f_1: \cZ \times \cR \rightarrow \cX$, $g: \cZ \times \bar{\cZ} \rightarrow \cK$, $g_1: \cX \times \bar{\cZ} \times \bar{\cR}  \rightarrow \cK$,
then, $K$ is a function of $X$ and $\bar{Z}$.
\end{claim}

\begin{proof}
The derivation below shows that $K$ is a function of $X$ and $\bar{Z}$, i.e., that $H(K) =I(K;X,\bar{Z})$.
\begin{align*}
H(K) 
& = 
I(K;X,\bar{Z},\bar{R}) \\
& =
I(K;X,\bar{Z})+I(K;\bar{R} \mid X,\bar{Z}) \\
& \leq 
I(K;X,\bar{Z})+I(K,X,\bar{Z};\bar{R}) \\
& \leq
I(K;X,\bar{Z})+I(Z,R,\bar{Z};\bar{R})\\
& =
I(K;X,\bar{Z}) \leq H(K).
\end{align*}
\end{proof}

\begin{claim}
\label{claim:func}
Let $Z_1,Z_2,Z_{3},R$ be independent random variables.
Let $X_1=f_1(Z_1,R)$, and $X_2=f_2(Z_2,R)$ for functions $f_1: \cZ_1 \times \cR \rightarrow \cX_1$ and $f_2: \cZ_2 \times \cR \rightarrow \cX_2$, respectively.
Suppose that $K$ can be recovered in three ways as 
\begin{itemize}
\item $K=g(Z_1,Z_2,Z_{3})$, 
\item $K= g_1(Z_1,X_2,Z_{3})$, and 
\item $K=g_2(X_1,Z_2, Z_{3})$,
\end{itemize}
for functions $g: \cZ_1 \times \cZ_2 \times \cZ_3 \rightarrow \cK$, $g_1: \cZ_1 \times \cX_2 \times \cZ_3 \rightarrow \cK$, and $g_2: \cX_1 \times \cZ_2 \times \cZ_3 \rightarrow \cK$, respectively.
Then there exists a function $h: \cX_1 \times \cX_2 \times \cZ_3 \rightarrow \cK$ such that $K=h(X_1,X_2,Z_{3})$. 
\end{claim}

\begin{proof}
Consider the random variables $Z_1, Z_2,Z_{3}$, and $R$.
We would like to show the existence of $h: \cX_1 \times \cX_2 \times \cZ_3 \rightarrow \cK$ that determines $K$.
Define $h(x_1,x_2,z_{3})$ to be the set of values $k$ for which there exists $z_1,z_2,r$ such that $f_1(z_1,r)=x_1$, $f_2(z_2,r)=x_2$, and $g(z_1,z_2,z_{3})=k$.
In what follows, we show that $h(x_1,x_2,z_3)$ has size at most 1, implying that $h$ is a (partial) function.
This suffices for our proof.
For any $z_1,z_2,z_3,r$ and $k$, if $g(z_1,z_2,z_3)=k$, $x_1=f_1(z_1,r)$, and $x_2= f_2(z_2,r)$, it holds by our definitions that  $k$ is the only member in $h(x_1,x_2,z_3)$, implying that $h$ can recover $k$.

It is left to show that $h(x_1,x_2,z_3)$ is a (partial) function. 
We strongly rely on the fact that $K=g_1(Z_1,X_2,Z_{3})$ and that $K=g_2(X_1,Z_2,Z_{3})$. 
Specifically, for any $(z_1,z_2,z_{3},r)$, it holds that $g(z_1,z_2,z_{3})=g_1(z_1,f_2(z_2,r),z_{3})=g_2(f_1(z_1,r),z_2,z_{3})$.
To show that $h$ is a function, assume for $x_1,x_2,z_{3}$ that $(z_1,z_2,r)$ satisfies $f_1(z_1,r)=x_1$, $f_2(z_2,r)=x_2$, $g(z_1,z_2,z_{3})=k$, and that $(z'_1,z'_2,r')$ satisfies $f_1(z'_1,r')=x_1$, $f_2(z'_2,r')=x_2$, $g(z'_1,z'_2,z_{3})=k'$; we show that $k=k'$ by noticing that
\begin{align*}
k=g(z_1,z_2,z_{3})
& =
g_1(z_1,f_2(z_2,r),z_{3})=g_1(z_1,x_2,z_{3})\\
& =
g_1(z_1,f_2(z'_2,r'),z_{3})=g(z_1,z'_2,z_{3})\\
& =
g_2(f_1(z_1,r),z'_2,z_{3})= g_2(x_1,z'_2,z_{3})\\
& =
g_2(f_1(z'_1,r'),z'_2,z_{3})=g(z'_1,z'_2,z_{3}) \\
& = k'.
\end{align*}
\end{proof}

\subsection{Proof of Theorem~\ref{the:q1c}}
We are now ready to prove Theorem~\ref{the:q1c}.

\begin{proof} (of Theorem~\ref{the:q1c})
It suffices to prove only one direction of the asserted claim; the other direction is immediate.
We use the claims and notation introduced in Section~\ref{sec:pre}.
Consider a positive-rate key-cast scheme $({\mathcal F},\mathcal{G})=(\{f_{e}\},\{g_{i}\})$ for $\cI$, and let $S={\tt Support}(K)=\{s_1,\dots,s_k\}$.
Consider running Protocol~\ref{protocol:2} given in Section~\ref{sec:modify} for every $(s_i,d_j)$ pair in $S \times D$ (with independent and uniformly distributed message bit $m_i$ for source $s_i$).
After running the protocol for all $(s_i,d_j) \in S \times D$, each terminal can compute a new key $K^* = \sum_{i=1}^km_i$.
Note that for $(s_i,d_j) \in S \times D$, in the protocol corresponding to $(s_i,d_j)$ the only vertices in $V \setminus \{s_i,d_j\}$ that may learn information about $m_i$ are the unprotected cut vertices. 
Therefore, after running Protocol~\ref{protocol:2}, for every $(s_i,d_j)$ pair in $S \times D$, the only network nodes in $V \setminus S$ that may gain information about $K^*$ are those that are unprotected cut vertices for $S$.
Similarly, the only network nodes in $S$ that may gain information about $K^*$ are those $s_i \in S$ that are themselves unprotected cut vertices for a collection of pairs $\{(s_j,d_{\ell_j})\}_{j \ne i}$.
Thus, to prove that the protocol is secure, it suffices to show that the following statements hold.
\begin{itemize}
\item No vertex $u \in V \setminus S$ is an unprotected cut-vertex for $S$; the statement formulated in Corollary~\ref{cor:q1c}, which asserts that for any node $u \in V \setminus S$, the set 
\begin{align*}
S_u\triangleq\{s \in S \mid \exists \ d \in D\ \text{s.t.} \ \text{$u$ is an unprotected cut vertex} \ &\text{w.r.t. $(s,d)$}\}
\end{align*}
does not equal $S$, is equivalent.
\item For every $s_i \in S$, 
the set 
\begin{align*}
S_{s_i}=\{s \in S \setminus\{s_i\} \mid \exists \ d \in D\ \text{s.t.} \ \text{$s_i$ is an unprotected cut vertex} \ &\text{w.r.t. $(s,d)$}\} 
\end{align*}
does not equal $S \setminus \{s_i\}$.
\end{itemize}
We start by proving the former by contradiction. We then address the latter by reducing it to the former analysis.


Assume in contradiction that there exists a vertex $u$ which is an unprotected cut-vertex for $S$.
Let $\{(s_i,d_i)\}_{i=1}^{k}$ be the collection of pairs such that $u$ is an unprotected cut-vertex for the pair $(s_i,d_i)$.
We show that this contradicts the assumed existence of the original secure key-cast scheme $({\mathcal F},\mathcal{G})=(\{f_{e}\},\{g_{i}\})$ in which $K$ depends only on information generated at $S$.
Specifically, we show that $K$ can be determined at $u$ by its incoming information.

We use the following additional notation (building on the notation presented in Section~\ref{sec:pre}).
Consider an ordering (that will be presented shortly) on the sources in $S=\{s_1,\dots,s_k\}$.
For $s_i \in S$, let $S[i]=\cup_{j=1}^iS_j$ and $\bar{S}[i]=S \setminus S[i]$, where $S_i$ is the set of source nodes in $W_{s_i,u}$; notice thus that $(S[i],\bar{S}[i])$ is a partition of $S$, and, as $i$ increases, the size of $S[i]$ grows until eventually $S[k]=S$.
Similarly, let $Z[i]=(Z_j: j \in [i])$ and $\bar{Z}[i]=Z[k] \setminus Z[i]$; random variable $Z[i]$ represents the information generated by nodes in $S[i]$, where $Z[k]$ includes all information generated by $S$.
Finally, let $X[i]=(X_j: j\in[i])$; $X_i$ represents the information entering $u$ from nodes in $W_{s_i,u}$, and $X[i]$ represents the corresponding aggregated information $X_j$ for $j=1,\dots, i$.
By Claims~\ref{claim:wpath1} and \ref{claim:wpath}, for any $i \ne j$ the sets $S_i$ and $S_j$ are either equal or disjoint.
By Claim~\ref{claim:wpath}, for any $i$, if $s_i \not \in S[i-1]$ (implying that $S_i \ne S_j$ for $j<i$) then $S[i-1]\cap S_i= \phi$.
We show by induction on $i \in [k]$ that as long as $S[i-1]\ne S$, $K$ is a function of $X[i]$ and $\bar{Z}[i]$.

For the base case, consider the pair $(s_1,d_1)$, i.e., the case $i=1$.
First note that $S[0] = \phi$ and thus for $i=1$, $S[i-1]\ne S$.
By Claim~\ref{claim:sum} and Claim~\ref{claim:func0}, 
$K$ is a function of $X_1$ and $\bar{Z}_1$. 
When $i=1$, $X[1]=X_1$ and $\bar{Z}[1]=\bar{Z}_1$. 
We thus conclude the inductive hypotheses for $i=1$. Namely, that $K$ is a function of $X[1]$ and $\bar{Z}[1]$.
If $S=S[1]$ (that is, $\bar{S}[1]=\phi$) we reach a contradiction by noticing that no sources contribute to $\bar{Z}[1]$ and thus $H(K) =I(K;X[1])$, implying that node $u$ can recover $K$.
If, $S[1] \ne S$, then we continue in the inductive process.

Assume for $i-1$ that $S[i-1]\ne S$ (implying that  $\bar{S}[i-1] \ne \phi$), and that $K$ is a function of $X[i-1]$ and $\bar{Z}[i-1]$.
Set $s_i$ to be a vertex in $\bar{S}[i-1]$.
By Claim~\ref{claim:wpath}, $S[i-1]\cap S_i= \phi$.
Thus $\bar{Z}[i-1]$ is exactly $(Z_i,\bar{Z}[i])$.
So, by induction, $K$ is a function of $X[i-1]$, $Z_i$, and $\bar{Z}[i]$.
Moreover, by Claim~\ref{claim:sum} and Claim~\ref{claim:func0}, $K$ is a function of $X_i,\bar{Z}_i$.
Notice that $Z[k]=(Z[i-1],Z_i,\bar{Z}[i])$.
Thus, $\bar{Z}_i= (Z[i-1],\bar{Z}[i])$ and $K$ is a function of $X_i,Z[i-1]$, and $\bar{Z}[i]$.
Finally, as $S={\tt Support}(K)$, it holds that $K$ is a function of $(Z[i-1],Z_i,\bar{Z}[i])$.
Here, Claim~\ref{claim:func} implies that $K$ is a function of $X[i-1]$, $X_i$, and $\bar{Z}[i]$, which, in turn, implies the inductive hypothesis that $K$ is a function of $X[i]$ and $\bar{Z}[i]$.
Specifically, we apply Claim~\ref{claim:func} with $Z_1=Z[i-1]$, $Z_2=Z_i$, $Z_{3}=\bar{Z}[i]$, $X_1=X[i-1]$, $X_2=X_i$, and a random vector $R$ that consists of all randomness generated at nodes $v \not\in S$.

We conclude the inductive process once $S=S[i]$ (that is, $\bar{S}[i]=\phi$).
In this case, we reach a contradiction as no sources contribute to $\bar{Z}[i]$ and thus $H(K) =I(K;X[i])$, implying that node $u$ can recover $K$.

This concludes the proof of the first assertion that no vertex $u \in V \setminus S$ can be an unprotected cut-vertex for $S$.
More specifically, we have shown that in 
the positive-rate key-cast scheme $({\mathcal F},\mathcal{G})=(\{f_{e}\},\{g_{i}\})$ for $\cI=(G,V,D,\cB)$ in which $\cB=\{\beta_v \mid v \in V \setminus D, \beta_v=\text{In-edges}(v)\}$, setting $S={\tt Support}(K)=\{s_1,\dots,s_k\}$ guarantees that $S_u \ne S$ for every $u \in V \setminus S$.

We now address the second assertion.
Assume in contradiction that for some $s_i \in S$, 
the set $S_{s_i}$ defined previously equals $S \setminus \{s_i\}$.
To reach a contradiction, we first slightly modify the instance $\cI$ and the assumed positive-rate secure key-cast scheme $({\mathcal F},\mathcal{G})$.
Consider the instance $\hat{\cI}= (\hat{G},\hat{V},\hat{D},\hat{\cB})$ in which $\hat{G}=(\hat{V},\hat{E})$ is identical to  $G=(V,E)$ except for the addition of a single vertex $\hat{s}_i$ connected by an edge to the original $s_i \in S$.
We set $\hat{D}=D$ and $\hat{\cB}=\{\beta_v \mid v \in \hat{V} \setminus \hat{D}, \beta_v=\text{In-edges}(v)\}$.
Moreover, consider the modified key-cast scheme $(\hat{{\mathcal F}},\hat{\mathcal{G}})=(\{\hat{f}_{e}\},\{\hat{g}_{i}\})$ which is identical to the original  scheme $({\mathcal F},\mathcal{G})=(\{f_{e}\},\{g_{i}\})$ with the exception that random bits $M_{s_i}$ used by $s_i$ in the original scheme are no longer generated in the modified scheme $(\hat{{\mathcal F}},\hat{\mathcal{G}})$ at $s_i$ but rather are generated at $\hat{s}_i$ and transmitted along the edge $(\hat{s}_i,s_i)$.
It follows that: (i) applying code $(\hat{{\mathcal F}},\hat{\mathcal{G}})$ on the modified instance $\hat{\cI}$, terminals $\hat{D}=D$ decode the exact same key $K$ as in the original instance $\cI$ under code $({\mathcal F},\mathcal{G})$; (ii) in the modified setting, $\hat{S} \triangleq {\tt Support}(K)=S \cap \{\hat{s}_i\} \setminus \{s_i\}$; and (iii) $s_i \in \hat{V} \setminus \hat{S}$ is an unprotected cut vertex for $\hat{S}$ in $\hat{G}$.
However, by our proof of the first assertion (replacing $\cI$ and $({\mathcal F},\mathcal{G})$ by $\hat{\cI}$ and $(\hat{{\mathcal F}},\hat{\mathcal{G}})$),  $s_i$ can recover $K$ in $(\hat{{\mathcal F}},\hat{\mathcal{G}})$, which in turn implies that $s_i$ can recover $K$ in $({\mathcal F},\mathcal{G})$; this is the desired  contradiction to the assumed security of $({\mathcal F},\mathcal{G})$. This concludes our proof.
\end{proof}

\section{Conclusions and open problems}
\label{sec:conclude}

In this work, we study positive-rate secure multicast and positive-rate key-cast under the requirement that no single non-terminal network node can gain information about the shared key.
Using the notion of alternating paths, for both problems we present a combinatorial characterization of instances $\cI$ for which positive-rate is possible and present corresponding communication schemes over $GF_2$.
The analysis here does not address the (provably challenging) problem of optimizing the secret-message/key rate, a problem left to future studies.

\appendix


\subsection{Multiple terminal setting ($|D|>1$)}
\label{app:inter}
When $D$ includes more than a single terminal node, 
we take advantage of the fact that we are only interested in  characterizing positive rate and perform the scheme of Theorem~\ref{the:q1} for each terminal node $d \in D$ in parallel,
with independent randomness.
That is, for each $d \in D$, one uses independent uniform bits for the executions of Protocol~\ref{def:protocol} and for the variables $\{r_i\}$.
Specifically, let $M$ be the random variable corresponding to the source message $m$.
Let ${\tt Vars}_i$ be the random variables used in the scheme corresponding to $d_i$, and, for $v \in V$, let $X^{(i)}_{\tt In}(v)$ be the incoming information to $v$ in the scheme corresponding to $d_i$. 
Our communication scheme implies that $X^{(i)}_{\tt In}(v)$ consists of linear combinations of (mutually independent uniform) random variables taken from the set $\{M\} \cup {\tt Vars}_i$.
By Theorem~\ref{the:q1} it holds for each $i$ that $I(M;X^{(i)}_{\tt In}(v))=0$.
Moreover, for $i \ne i'$ the collection of variables ${\tt Vars}_i$ is disjoint and thus independent of ${\tt Vars}_{i'}$.
It follows that the total incoming information to vertex $v$, $X_{\tt In}(v)=(X^{(1)}_{\tt In}(v),\dots,X^{(|D|)}_{\tt In}(v))$ satisfies $I(M;X_{\tt In}(v))=0$.

Formally, consider the $\{0,1\}$ matrix $A$ tying $X_{\tt In}(v)$ and the variables in $(M,{\tt Vars}_1,\dots,{\tt Vars}_{|D|})$, {\em i.e.},  $X_{\tt In}(v)$ is obtained by the product of $A$ and the vector of random variables $(M,{\tt Vars}_1,\dots,{\tt Vars}_{|D|})$.
Let $A_i$ be the collection of rows in $A$ with non-zero entries corresponding to variables in ${\tt Vars}_i$.
The discussion above implies that $A_1,\dots,A_{|D|}$ is a partition of the rows in $A$.
Assume in contradiction that $I(M;X_{\tt In}(v)) \ne 0$.
Then the vector $(1,0,\dots,0)$ (with a value of 1 in the entry corresponding to the variable $M$) is in the row-span of $A$. 
This implies the existence of vectors $a_1,\dots, a_{|D|}$, with $a_i$ in the row-span of $A_i$, such that $\sum_{i=1}^{|D|} a_i=(1,0,\dots,0)$; which, in turn, implies, by the definition of $A_1,\dots,A_{|D|}$, that at least one vector $a_i$ in the collection $a_1,\dots, a_{|D|}$, must equal $(1,0,\dots,0)$. 
This contradicts our prior observation under Theorem~\ref{the:q1} that $I(M;X^{(i)}_{\tt In}(v))=0$.

\subsection{Proof of Theorem~\ref{the:q1b} (stated below)}
\label{app:keycast}

\begin{theorem}
\label{the:q1b}
Let $\cI=(G,V,D,\cB)$ with $\cB=\{\beta_v \mid v \in V \setminus D, \beta_v=\text{In-edges}(v)\}$.
Then $\cI$ has key-cast rate $R_\rk>0$ according to Definition~\ref{def:key} if (i) there exists a node $s$ such that, for all $d\in D$, every cut vertex $u$ separating the pair $(s,d)$ is protected, and, in addition, (ii) there exists a node $s'$ such that, for all $d \in D$, there is a path from $s'$ to $d$ that does not pass through $s$.
\end{theorem}

\begin{proof}
The key dissemination protocol proceeds as follows.
Using Theorem~\ref{the:q1} and Corollary~\ref{cor:q1}, source node $s$ securely multicasts a (uniformly distributed) message bit $m$ to all terminals in $D$. 
In parallel, source node $s'$ sends a (uniformly distributed) bit $m'$ to all terminal nodes in $D$ along the designated paths (which do not pass through $s$), where $m'$ is independent of all information used in the secure transmission of $m$.
Below, denote the random variables corresponding to $m$ and $m'$ by $M$ and $M'$ respectively.
The key $K$ is set to $m+m'$.

For the security analysis, let $v$ be a node in $G$.
Vector $X_{{\rm In}(v)}$ consists of at most two independent parts, one corresponding to the secure transmission of $m$, denoted $X^{(m)}_{{\rm In}(v)}$, and the second corresponding to the transmission of $m'$, denoted $X^{(m')}_{{\rm In}(v)}$.
When node $v$ is not used in the transmission of $m$, we set $X^{(m)}_{{\rm In}(v)}$ to be deterministically equal to 0.
When node $v$ is used in the transmission of $m'$, it holds that $X^{(m')}_{{\rm In}(v)}=m'$,  otherwise we set $X^{(m')}_{{\rm In}(v)}$ to be deterministically equal to 0.
By the security of the transmission of $m$, it holds for any node $v \not \in D \cup \{s\}$ that $I(M,M'; X^{(m)}_{{\rm In}(v)})=0$.

Thus for $v \not \in D \cup \{s\}$, 
\begin{align*}
I(K; X_{{\rm In}(v)})& = I(M+M'; X^{(m)}_{{\rm In}(v)},X^{(m')}_{{\rm In}(v)})\\
& \leq 
I(M+M'; X^{(m)}_{{\rm In}(v)},M')\\
& = 
I(M+M'; M')+ I(M+M'; X^{(m)}_{{\rm In}(v)} | M') \\
& \leq
0+I(M+M',M'; X^{(m)}_{{\rm In}(v)})=0.
\end{align*}

For $s$ it holds that $I(M'; X_{{\rm In}(s)})=0$ and thus $I(K; X_{{\rm In}(s)})=0$. 
\end{proof}
 
%

\bibliographystyle{unsrt}
\bibliography{ref}
\end{document}